\documentclass[prb,superscriptaddress,twocolumn]{revtex4}
\usepackage[utf8]{inputenc}
\usepackage{amsmath}
\usepackage{graphicx}
\usepackage{multirow}
\usepackage{color}

\begin{document}
\definecolor{orange}{rgb}{1, 0.5, 0.0}
\title{Reciprocal-space implementation of rotational invariance for first-principles phonons and application to low-dimensional materials}

\author{Benoit Van Troeye}
\affiliation{Imec, Leuven, Belgium}
\author{Xavier Gonze}
\affiliation{Institute of Condensed Matter and Nanosciences, Universit\'{e} catholique de Louvain, Chemin des étoiles 8, B-1348 Louvain-la-Neuve, Belgium}
\author{Geoffrey Pourtois}
\affiliation{Imec, Leuven, Belgium}

\begin{abstract}
   The acoustic flexural (phonon) modes of low-dimensional materials should show a quadratic dispersion close to the Brillouin zone center. Any departure from this behavior in Density Functional Theory calculations is typically associated with a breaking of the rotational invariance, with few methods available to correct it. In this work, we reexamine this issue based on a reciprocal-space imposition of this condition, with corrections on the zone-center IFCs and their first and second derivatives with respect to the phonon wavevector. 
   We propose two correction schemes for the short-range part of the interatomic force constants, one based on the Moore-Penrose pseudoinverse, that we implement in the ABINIT software package, and one based on the straight modification of the on-site antisymmetric part of the first-order derivative of the dynamical matrix with respect to the wavevector. 
   We investigate the impact of imposing rotational invariance for different system dimensionalities and pseudopotentials. Finally, we discuss rotational invariance in the context of long-range electrostatics contribution to the IFCs. We observe that the usual treatment of electrostatics is by construction not rotationally invariant. This is found especially critical when correcting the second derivatives of the IFCs.
\end{abstract}
\maketitle

Due to their peculiar physics compared to bulk materials~\cite{Novoselov2016} and the on-going progresses towards their experimental synthesis and industrial applications~\cite{Lemme2022,Obrien2023}, low-dimensional materials have attracted a steady interest from the scientific community through the years. 
First-principles computations for these materials have successfully predicted their properties for a variety of systems~\cite{Lin2023}. 
However, a series of technicalities arising from their confined nature needs to be tackled in order to untangle real physics from the one originating from implementation and numeric. 
Indeed, most of the first principles implementations for condensed matter work with periodic boundaries conditions along the three Cartesian directions, while the confinement directions require specific treatments to remove the spurious Coulomb interaction between the periodic images~\cite{Sohier2017b}. 
The proper description of the long-range decay of the interatomic force constants (IFCs) in 2D materials, arising from dipole-dipole and further-order electrostatic interactions, has only been investigated recently~\cite{Sohier2016,Sohier2017,Royo2021}.
Meanwhile, the existence of a mode with quadratic dispersion in 2D material has been suggested in the first work of 
Landau
based on elastic wave considerations~\cite{Landau2012}, but such quadratic
behavior is difficult to obtain from usual first-principles simulations.
In the continuation of this premise, different approaches have been developed over the years to correct first-principles data and to recover this quadratic behavior, with varying degrees of success~\cite{Carrete2016,Lin2022}.

The difficulty to obtain a quadratic dispersion is typically associated with the breaking of the IFCs rotational invariance, that states that a global rotation of the system should leave the energy invariant, and consequently generates no force nor torque on the atoms. 
With usual Density Functional Theory 
(DFT) approximations, this invariance, as well as the translational invariance (acoustic sum rule)~\cite{Gonze1997}, are typically broken and need to be restored afterwards; currently, only the acoustic sum rule is routinely corrected in modern DFT codes, with the exception of Quantum Espresso that provides a real-space implementation of the rotational invariance~\cite{Lin2022}. 
This question of linear versus quadratic phonon dispersion of the flexural mode is particularly important for thermal conductivity, since the product of the density of states and of Bose-Einstein distribution diverges for the quadratic case at low energy whatever the temperature. 

In this work, we revisit this question of rotational invariance and introduce an alternative approach that enforces it in reciprocal space, by means of the Fourier transforms of IFCs and, notably, their derivatives at the zone center. 
We then explore its impact on systems with different dimensionalities. 
Consistently with previous reports~\cite{Croy2020,Lin2023}, we find that the flexural acoustic modes of low-dimensional materials should show a quadratic dispersion close to the zone center, both following a perturbation theory approach and an analytical model of atoms bound by springs. 
Rotational invariance enforces conditions on the first and second derivatives of the IFCs with respect to the phonon wavevector, as well as on the zone-center IFCs. 
Such derivatives are readily available using the long-wavelength implementation in the ABINIT software~\cite{Stengel2013,Royo2019,Royo2022}. 
We show that they can be corrected alongside the zone-center IFCs following two proposed method : a pseudo-inverse approach and a direct imposition on the derivatives of the IFCs with respect to the phonon wavevector. 
Some challenges remain however with the long-range electrostatics contributions to the IFCs (dipole-dipole~\cite{Gonze1997}, dipole-quadrupole~\cite{Stengel2013,Royo2019,Royo2022}, ...), whose implementations in ab initio software is not rotationally invariant for materials with anisotropic dielectric tensor. This especially causes problems when enforcing rotational invariance on the second derivatives of the IFCs, which we are thus currently not correcting.
We report the implementation of these theoretical developments in the ABINIT~\cite{Abinit2005,Abinit2009,Gonze2016} software package and apply the methodology on a few selected systems (phosphorene, WS$_2$, H$_2$O, CO, polyethylene, graphite).

\section{Theory} \label{sec:theory}

In order to properly introduce the rotational invariance of the interatomic force constants, let us first recall the translational invariance condition:
\begin{equation}
\sum_{\kappa'\mathbf{b}} \Phi_{\kappa\alpha,\kappa'\beta}(\mathbf{0},\mathbf{b}) = \sum_{\kappa'}\tilde{\Phi}_{\kappa\alpha,\kappa'\beta}(\mathbf{0})= 0.~\label{eq:transInv}
\end{equation}
In this equation, $\Phi_{\kappa\alpha,\kappa'\beta}(\mathbf{0},\mathbf{b})$ is the interatomic force constant (second derivative of the energy with respect to nuclei displacements, or equivalently, minus the first derivative of the force on one nucleus when another nucleus is displaced) between the nuclei $\kappa$ in cell $\mathbf{0}$ and $\kappa'$ in cell $\mathbf{b}$, $\alpha,\beta$ being Cartesian directions. 
The summation on $\kappa'$ runs from 1 to $N_{\textrm{at}}$, the number of atoms in the primitive cell. For sake of simplicity, the latter is not explicitly mentioned in the summation on $\kappa'$ in Eq.~(\ref{eq:transInv}), or, later, in similar summations on the atoms in the primitive cell with dummy index $\kappa$.

The second quantity in Eq.~(\ref{eq:transInv}) is its Fourier transform,
\begin{multline}
\tilde{\Phi}_{\kappa\alpha,\kappa'\beta}(\mathbf{q})= \frac{1}{N}\sum_{\mathbf{a},\mathbf{b}} \Phi_{\kappa\alpha,\kappa'\beta}(\mathbf{a},\mathbf{b}) e^{2\pi j\mathbf{q}.\left(\mathbf{R}^\mathbf{b}_{\kappa'}-\mathbf{R}^\mathbf{a}_{\kappa}\right)} \\ =\sum_{\mathbf{b}} \Phi_{\kappa\alpha,\kappa'\beta}(\mathbf{0},\mathbf{b}) e^{2\pi j\mathbf{q}.\left(\mathbf{R}^\mathbf{b}_{\kappa'}-\mathbf{R}^\mathbf{0}_{\kappa}\right)};\\ \label{eq:ifc}
\end{multline}
where $\mathbf{R}^{\mathbf{b}}_{\kappa}$ corresponds to the Cartesian coordinates of the nucleus $\kappa$ in cell $\mathbf{b}$, $\mathbf{q}$ the phonon wavevector and $N$ is the number of considered unit cells in the summation (Born-von Karman supercell). Note that this convention is slightly different than in Ref.~\onlinecite{Gonze1997} due to a phase factor, which renders it more convenient for rotational invariance understanding. Then, the diagonalization of this matrix, normalized by the nuclei mass $M_\kappa$, gives the square of the phonon frequencies $\omega^2_{m\mathbf{q}}$ as well as the associated eigenvectors $U_{m\mathbf{q}}(\kappa\alpha)$ ($m$ is the phonon branch index):
\begin{equation}
\sum_{\kappa'\beta}\tilde{\Phi}_{\kappa\alpha,\kappa'\beta}(\mathbf{q})U_{m\mathbf{q}}(\kappa'\beta)=M_\kappa\omega^2_{m\mathbf{q}}U_{m\mathbf{q}}(\kappa\alpha). \label{eq:dynmat}
\end{equation}

Eq.~(\ref{eq:transInv}) enforces that the total energy of the system is invariant upon global translations. This guarantees that three zero-frequency modes are present at the zone center, independently of the problem dimensionality (0D $\rightarrow$ 3D). 
These modes belong to the acoustic phonon branches, hence the condition Eq.~(\ref{eq:transInv})
 is also named Acoustic Sum Rule (ASR).

In many implementations of DFT, translational invariance is slightly broken due to the evaluation of the exchange-correlation energy on a finite real-space grid, as well as due to the non-linear core corrections of the pseudopotentials (if any). 
Recent generations of pseudopotentials are now explicitly taking the minimization of the acoustic sum rule breaking as a metric to assess pseudopotential quality~\cite{pseudodojo2018}.
Translational invariance can however be easily corrected, for example by adjusting the diagonal on-site IFCs~\cite{Gonze1997}, a standard procedure in most DFT codes.

Meanwhile, the imposition of rotational invariance, i.e. the invariance of the energy
upon a global rotation of the system, is much less frequent. 
This invariance of energy also
implies the cancellation of
the total torque on a nucleus when all atoms are moved away infinitesimally from their relaxed position in a way that mimics a global rotation of the system around this atom. 
Rotation invariance imposes conditions on the first moment of real-space IFCs~\cite{Pick1970}
\begin{equation}
\sum_{\beta\gamma}
\sum_{\kappa'\mathbf{b}} \epsilon_{\beta\gamma\delta}\Phi_{\kappa\alpha,\kappa'\beta}(\mathbf{0},\mathbf{b}) R^\mathbf{b}_{\kappa'\gamma}= 0,~\label{eq:rotInv}
\end{equation}
 where $\delta,\gamma$ are Cartesian directions and $\epsilon_{\beta\gamma\delta}$ is the Levi-Civita tensor. In principle, it is possible to brute-force impose this condition on the real-space IFCs using for example a least-squares method (see Ref.~\onlinecite{Lin2022}). 
 However, such a method, due to the presence of the $R^\mathbf{b}_{\kappa'\gamma}$ factor in the condition matrix, will propagate corrections mostly to the farthest IFCs. This typically leads to non-analytic behavior in reciprocal space and can lead to numerical instabilities that we discuss in more detail in App.~A. 
 Hereafter, we explore another way to impose these conditions, based on a reciprocal space expression.

Indeed, 
combining this constraint with the translational invariance, 
Eq.~(\ref{eq:transInv}), gives
\begin{equation}
\sum_{\beta\gamma}\sum_{\kappa'\mathbf{b}} \epsilon_{\beta\gamma\delta}\Phi_{\kappa\alpha,\kappa'\beta}(\mathbf{0},\mathbf{b}) (R^\mathbf{b}_{\kappa'\gamma}-R^\mathbf{0}_{\kappa\gamma}) = 0.~\label{eq:rotInv2}
\end{equation}
Then, 
similarly to Ref.~\onlinecite{Royo2022}, in the 3D case, we note that it is possible to express the rotational invariance condition in reciprocal space by taking the derivative of Eq.~(\ref{eq:ifc}) w.r. to the phonon wavevector and estimating it at the zone center:
\begin{equation}
 \sum_{\beta\gamma}\sum_{\kappa'}
 \epsilon_{\beta\gamma\delta} \left. 
 \frac{\partial\tilde{\Phi}_{\kappa\alpha,\kappa'\beta}}
 {\partial q_{\gamma}}\right|_{\mathbf{q}=\mathbf{0}} \triangleq \sum_{\kappa'} \sum_{\beta\gamma}2\pi\tilde{\Phi}^{(1)}_{\kappa\alpha,\kappa'\beta,\gamma}\epsilon_{\beta\gamma\delta}= 0, \label{eq:rot_3d}
 \end{equation}
or equivalently, 
$[ \vec{\nabla}_{\mathbf{q}} \times \sum_{\kappa'} \vec{\Phi}_{\kappa \alpha,\kappa'} ]|_{\mathbf{q}=0}=0$,
where the vector $\vec{\Phi}_{\kappa \alpha,\kappa'} $ has components
$\tilde{\Phi}_{\kappa\alpha,\kappa'\beta}$. The notation $\tilde{\Phi}^{(1)}_{\kappa\alpha,\kappa'\beta,\gamma}$ for the IFCs first derivatives with respect to the wavevector is introduced here for sake of brevity. 
The combined ASR,
Eq.~(\ref{eq:transInv}),
and rotational invariance conditions, 
Eq.~(\ref{eq:rot_3d}), imposed at the same time, will be denoted as 
acoustic and moment sum rules (AMSR) in the following.

Instead, if one considers that the system is confined along $z$ (2D case), then $q_z$ is not defined, while $R^\mathbf{b}_{\kappa z}=R^\mathbf{0}_{\kappa z}$ by translational invariance (specifically for the $z$ coordinate). Then, the rotational invariance in reciprocal space becomes:
\begin{equation}
\left \{
   \begin{array}{r c l}
     T_{\kappa\alpha,x} - \sum_{\kappa'}\left. \Im(\tilde{\Phi}^{(1)}_{\kappa\alpha,\kappa'z,y})\right|_{\mathbf{q}=\mathbf{0}} & &=0 \\
        T_{\kappa\alpha,y} +   \sum_{\kappa'}\left. \Im(\tilde{\Phi}^{(1)}_{\kappa\alpha,\kappa'z,x})\right|_{\mathbf{q}=\mathbf{0}}& &=0 \\
       \sum_{\beta\gamma}\sum_{\kappa'}\left. 
       \epsilon_{\beta\gamma z}
       \Im(\tilde{\Phi}^{(1)}_{\kappa\alpha,\kappa'\beta,\gamma})\right|_{\mathbf{q}=\mathbf{0}} &&= 0 \\
   \end{array}
   \right . \label{eq:2dcase}
\end{equation}

with $\mathbf{q}=(q_x,q_y,0)$ purely in-plane, $\Im(f)$ the imaginary part of the $f$ function. We have introduced
\begin{align}
    T_{\kappa\alpha,\gamma} &= 
    \sum_{\kappa',\beta} 
    \epsilon_{\beta z\gamma}
    (R^\mathbf{0}_{\kappa'z}-R^\mathbf{0}_{\kappa z}) \tilde{\Phi}_{\kappa\alpha,\kappa'\beta} (\mathbf{0}) 
    \nonumber\\
    &\triangleq \sum_{\kappa'\beta} M^{(1)}_{\kappa\alpha,\kappa'\beta,z} \epsilon_{\beta z \gamma} \label{eq:torque},
\end{align}
the zone-center residual torque in the system, directly related to the real-space first moment of the IFCs $M^{(1)}_{\kappa\alpha,\kappa'\beta,\gamma}$. 
The latter quantity being real, one can also deduce that the $q_x$ and $q_y$ derivatives of the Fourier transform of the IFCs must be purely imaginary, which also naturally arise due to the Hermiticity of the reciprocal IFCs matrix. 

Eq.~(\ref{eq:2dcase}) imposes a set of conditions on the zone-center and the first-order IFCs derivatives. 
However, rotational invariance also propagates to the higher IFCs derivatives, for example to the second derivatives. 
Indeed,
using the periodicity of the lattice and after the application of rotational invariance on both sides on the real-space IFCs matrix, one obtains

{\small
\begin{align}
\sum_{\kappa\kappa'}\tilde{\Phi}^{(2)}_{\kappa z,\kappa' z,\gamma\delta}
&\triangleq \frac{1}{(2\pi)^2} \sum_{\kappa\kappa'}\left.\frac{\partial^2\tilde{\Phi}_{\kappa z,\kappa'z}}{\partial q_\gamma \partial q_\delta}\right|_0
\nonumber
\\
=\frac{-1}{N}\sum_{\kappa\mathbf{a},\kappa'\mathbf{b}}
&\Phi_{\kappa z,\kappa' z}(\mathbf{a},\mathbf{b})(R^\mathbf{b}_{\kappa'\gamma}-R^\mathbf{a}_{\kappa\gamma})(R^\mathbf{b}_{\kappa'\delta}-R^\mathbf{a}_{\kappa'\delta})
\nonumber
\\
=\frac{1}{N}\sum_{\kappa\mathbf{a},\kappa'\mathbf{b}}
&\Phi_{\kappa z,\kappa' z}(\mathbf{a},\mathbf{b}) [R^\mathbf{a}_{\kappa\gamma}R^\mathbf{b}_{\kappa'\delta}+R^\mathbf{a}_{\kappa\delta}R^\mathbf{b}_{\kappa'\gamma}] 
\nonumber
\\
=-\sum_{\kappa\kappa'}
&\tilde{\Phi}_{\kappa \gamma,\kappa'\delta}(\mathbf{0})(R^\mathbf{0}_{\kappa'z}-R^\mathbf{0}_{\kappa z})^2 \nonumber
\\
\triangleq -\sum_{\kappa\kappa'}
&
M^{(2)}_{\kappa\gamma,\kappa'\delta,zz}, \label{eq:d2dq}
\end{align}
}
We denote Eq.~(\ref{eq:d2dq}) as the second-order conditions of rotational invariance to differentiate it later from  Eq.~(\ref{eq:2dcase}), with the latter referred as the first-order conditions consequently.

Both constraints on the IFCs derivatives modify the dispersion of the acoustic phonon mode around the zone center. Applying perturbation theory and treating the phonon wavevector as the perturbation~\cite{Born1996}, one can actually show that the phonon dispersion is purely quadratic for flexural modes (see App.~B and App.~C), as already highlighted in Refs.~\onlinecite{Croy2020} and~\onlinecite{Lin2022}, with the linear term vanishing entirely because of translational and rotational invariance. Any variation from this behavior at the DFT level stems from the breaking of rotational invariance, either at the level of the first derivatives of the IFCs,
Eq.~(\ref{eq:2dcase}), or at the level of their second derivatives, Eq.~(\ref{eq:d2dq}). To restore it, one can play either on the zone-center IFCs or on the IFCs derivatives. We will discuss how to apply such a correction in the next section.

%

For the 1D case (periodic direction along $x$), a similar set of equations and consequences can be derived for rotational invariance:

\begin{equation}
\left \{
   \begin{array}{r c l}
      \sum_{\kappa'} \tilde{\Phi}_{\kappa\alpha,\kappa'y}(\mathbf{0}) \Delta R_{\kappa\kappa'z}- \tilde{\Phi}_{\kappa\alpha,\kappa'z}(\mathbf{0}) \Delta R_{\kappa\kappa'y}& &=0 \\
       \sum_{\kappa'} \tilde{\Phi}_{\kappa\alpha,\kappa'x}(\mathbf{0}) \Delta R_{\kappa\kappa'z} -    \Im( \tilde{\Phi}^{(1)}_{\kappa\alpha,\kappa'z,x})& &=0 \\
      -\sum_{\kappa'} \tilde{\Phi}_{\kappa\alpha,\kappa'x}(\mathbf{0}) \Delta R_{\kappa\kappa'y} +  \frac{1}{2\pi} \Im(\tilde{\Phi}^{(1)}_{\kappa\alpha,\kappa'y,x})& &=0 \\
       
   \end{array}
   \right . \label{eq:1dcase}
\end{equation}

with $\Delta R_{\kappa\kappa'\alpha} = R^0_{\kappa'\alpha}-R^0_{\kappa \alpha}$. In this case, one of the zone-center modes must cancel out to impose rotational invariance. The dispersion of the dynamical matrix eigenvalues is also impacted, similarly to the 2D case. 
\section{Implementation of rotational invariance}
\label{sec:implementation}

In real space, the imposition of rotational invariance on IFCs
cannot be disentangled from the imposition of the translational invariance, since both act directly on the same variable landscape, namely the IFCs of the system~\cite{Lin2022}. Symmetry conditions also enter in the game, so that the proposed treatment for the imposition of the rotational invariance is based on a least-square or
compressive sensing technique~\cite{Lin2022}, working globally on all IFCs independent degrees of freedom. 
However, the number of such independent degrees of freedom scales quite unfavorably with the number of atoms in the primitive cell and the extent of the IFCs. The CPU time needed to impose the different constraints scales even worse than the straight number of degrees of freedom.

Working in reciprocal space opens new possibilities, as we will demonstrate. Our procedure has two steps: first, imposing the conditions on the zone-center dynamical matrix as well as its 
wavevector derivatives, and then, in a second step, from such corrections obtain the corrections to be done to an arbitrary wavevector dynamical matrix.
In such a procedure, one does not have to compute real space IFCs, hence this procedure is insensitive to the range of the IFCs: the number of atoms to be taken into account is always the one of the primitive cell.

In the first step of this procedure, we define the variable vector
\begin{equation}
\label{eq:variable_vector}
\mathbf{x}=
\begin{pmatrix}
\Re(\tilde{\Phi}_{\kappa\alpha,\kappa'\beta}(\mathbf{0}))\\
\Im(\tilde{\Phi}^{(1)}_{\kappa\alpha,\kappa'\beta,\gamma}) \\
\Re(\tilde{\Phi}^{(2)}_{\kappa\alpha,\kappa'\beta,\gamma\delta}) \\
\end{pmatrix},
\end{equation}
where the entries are the components of the dynamical matrix at 
$\mathbf{q}=\mathbf{0}$, their first derivatives with respect to the wavevector, still at $\mathbf{q}=\mathbf{0}$, and their second derivatives with respect to the wavevector, also at $\mathbf{q}=\mathbf{0}$.
Note that the dynamical matrix at $\mathbf{q}=\mathbf{0}$
is purely real and symmetric (with respect to the joint exchange of $\kappa\alpha$ and 
$\kappa'\beta$ indices), as seen from 
Eq.~(\ref{eq:ifc}). Similarly, one deduces that its first
derivative with respect to the wavevector is purely imaginary, 
and antisymmetric for such exchange, and its second derivative
is purely real, and symmetric for such exchange. Moreover
the latter is symmetric with respect to the exchange of $\gamma$ and $\delta$ indices.

Taking into account these considerations for the compression of this vector, the right-hand side of 
Eq.~(\ref{eq:variable_vector}) is determined by
$3N_{\textrm{at}}(3N_{\textrm{at}}+1)/2$ real numbers for its first line, $3(3N_{\textrm{at}})(3N_{\textrm{at}}-1)/2$ real numbers
for its second line, and $6(3N_{\textrm{at}})(3N_{\textrm{at}}+1)/2$ real numbers
for its third line.

The condition matrix $\underline{\underline{A}}$, that is
such that
$\underline{\underline{A}}\mathbf{x}$ vanishes if $\mathbf{x}$ fulfills the conditions,
is constructed based on Eq.~(\ref{eq:transInv}), as well as the relevant equations among Eqs.~(\ref{eq:rot_3d}),~(\ref{eq:2dcase}),
~(\ref{eq:d2dq}) and~(\ref{eq:1dcase}), depending on the dimensionality of the problem.

The first possibility for imposing the constraints on the vector 
Eq.~(\ref{eq:variable_vector}) is similar in spirit  to the one followed for 
the real-space treatment~\cite{Lin2022}, albeit working in a smaller dimensionality space (the dimension of the vector in 
Eq.~(\ref{eq:variable_vector}) is smaller than the one of the set of IFCs due to its spatial extension).
Namely, one can resort to least-square or compressed-sensing techniques. 
Among these, the Moore-Penrose pseudoinverse approach has been used
for our implementation
in the ABINIT software package~\cite{Abinit2005,Abinit2009,Romero2020,Gonze2016}. The developments were performed in the ABINIT v10.6 and planned to be available by the end of the year 2026.

The translational and rotational invariances form the conditions of the matrix problem $\underline{\underline{A}}$, as well as the Hermiticity of the IFCs matrix.
Except for the case of one atom per cell, the space of variables is larger than the number of constraints, and the result is non-unique.
The pseudoinverse gives the solution in the least-square sense i.e. smallest change of the initial IFCs that satisfies the above-mentioned conditions. 
If $\mathbf{x}$ does not fulfill the conditions,
the correction to be added to $\mathbf{x}$ through the pseudo-inverse is 
\begin{equation}
\mathbf{\Delta}=  - \underline{\underline{A}}^+ \underline{\underline{A}} \mathbf{x},\label{eq:pseudoinverse}
\end{equation}
where $\underline{\underline{A}}^+$ is the pseudo-inverse of the condition matrix.

The second possibility for imposing the constraints on
the vector Eq.~(\ref{eq:variable_vector}) is
closer in spirit to the way the ASR is usually imposed
\cite{Gonze1997}.
Such imposition of the ASR proceeds as follows. From Eq.~(\ref{eq:transInv}), one replaces
the site-diagonal elements of the 
$\mathbf{q}=\mathbf{0}$ dynamical matrix computed from first principles (``old''):
\begin{equation}
\tilde{\Phi}_{\kappa\alpha,\kappa\beta}^{\mathrm{new}}(\mathbf{0})= 
\tilde{\Phi}_{\kappa\alpha,\kappa\beta}^{\mathrm{old}}(\mathbf{0})
-\sum_{\kappa''}
\tilde{\Phi}_{\kappa\alpha,\kappa''\beta}^{\mathrm{old}}(\mathbf{0})
.~\label{eq:ASRimposition}
\end{equation}
The new dynamical matrix at $\mathbf{q}=\mathbf{0}$ fulfills the ASR.
In other words, the correction of the dynamical matrix (``order zero'' derivative in the vector
Eq.~(\ref{eq:variable_vector}))
to fulfill the ASR
is
\begin{equation}
\Delta^{(0)}_{\kappa\alpha,\kappa\beta}= 
-\sum_{\kappa''}
\tilde{\Phi}_{\kappa\alpha,\kappa''\beta}^{\mathrm{old}}(\mathbf{0})
.~\label{eq:correction_ASR}
\end{equation}

We show how the rotational invariance can be imposed similarly in the three-dimensional case. 
Specifically, the rotational invariance constraint 
Eq.~(\ref{eq:rot_3d}) will be imposed to the second line of the vector 
Eq.~(\ref{eq:variable_vector}).
The first derivative of the dynamical matrix with respect to the wavevector
can be decomposed in its symmetric and anti-symmetric parts with respect to the exchange of its 
$\beta$ and $\gamma$ subscripts,
\begin{equation}
\label{eq:decompose_1}
\tilde{\Phi}^{(1)}_{\kappa\alpha,\kappa'\beta,\gamma}=
(\tilde{\Phi}^{(1)}
_{\textrm{symm}}
)
_{\kappa\alpha,\kappa'\beta,\gamma}
+
(\tilde{\Phi}^{(1)}
_{\textrm{anti}}
)
_{\kappa\alpha,\kappa'\beta,\gamma} \, ,
\end{equation}
with 
\begin{equation}
\label{eq:phi1_Symm}
(\tilde{\Phi}^{(1)}
_{\textrm{symm}}
)
_{\kappa\alpha,\kappa'\beta,\gamma}
=
\frac{1}{2}
\Big(
\tilde{\Phi}^{(1)}_{\kappa\alpha,\kappa'\beta,\gamma}
+
\tilde{\Phi}^{(1)}_{\kappa\alpha,\kappa'\gamma,\beta}
\Big)
\end{equation}
and
\begin{equation}
\label{eq:phi1_Anti}
(\tilde{\Phi}^{(1)}
_{\textrm{anti}}
)
_{\kappa\alpha,\kappa'\beta,\gamma}
=
\frac{1}{2}
\Big(
\tilde{\Phi}^{(1)}_{\kappa\alpha,\kappa'\beta,\gamma}
-
\tilde{\Phi}^{(1)}_{\kappa\alpha,\kappa'\gamma,\beta}
\Big).
\end{equation}
$\tilde{\Phi}^{(1)}_{\textrm{symm}}$ fulfills immediately the constraints 
Eq.~(\ref{eq:rot_3d}), hence one is left with their imposition on the antisymmetric part.
Next, we define an `axial' version of the antisymmetric part of the first-order derivative of the dynamical matrix, by converting the three possible pairs of 
$\beta\gamma$ subscripts (taking into account the antisymmetry) to the subscript 
$\delta$ of an axial vector:
\begin{equation}
\label{eq:phi_rotational}
(\tilde{\Phi}^{(1)}
_{\textrm{ax}}
)
_{\kappa\alpha,\kappa'\delta}
=
\sum_{\beta\gamma}
\epsilon_{\beta\gamma\delta}(\tilde{\Phi}^{(1)}_{\textrm{anti}}
)
_{\kappa\alpha,\kappa'\beta,\gamma}
\end{equation}
The knowledge of 
$(\tilde{\Phi}^{(1)}
_{\textrm{ax}}
)
_{\kappa\alpha,\kappa'\delta}
$,
combined with the knowledge of the symmetric part of the first-order derivative of the dynamical matrix
allows one to reconstruct the second line 
of Eq.~(\ref{eq:variable_vector}).

The rotational invariance constraints 
Eq.~(\ref{eq:rot_3d}) becomes simply
\begin{equation}
\sum_{\kappa'}
(
\tilde{\Phi}
_{\textrm{ax}}
)
_{\kappa\alpha,\kappa'\delta}
= 0.
~\label{eq:Rot_Imposition}
\end{equation}
The number of such constraints is 
$3.(3N_{\mathrm{at}})$.

These constraints can be imposed similarly to 
Eq.~(\ref{eq:ASRimposition}), 
by simply modifying the site-diagonal elements of the axial formulation:
\begin{equation}
(\tilde{\Phi}
_{\textrm{ax}}
^{\mathrm{new}}
)
_{\kappa\alpha,\kappa\delta}
=
(\tilde{\Phi}
_{\textrm{ax}}
^{\mathrm{old}}
)
_{\kappa\alpha,\kappa\delta}
-\sum_{\kappa''}
(
\tilde{\Phi}
_{\textrm{ax}}
^{\mathrm{old}})
_{\kappa\alpha,\kappa''\delta},
~\label{eq:rot_invar_imposition}
\end{equation}
Afterwards, the second line of 
Eq.~(\ref{eq:variable_vector}) can be reconstructed.
Equivalently,  
the correction of the $\kappa$-diagonal elements of the axial part of the first-order dynamical matrix 
is
\begin{equation}
(\Delta^{(1)}_\textrm{ax})_{\kappa\alpha,\kappa\beta}= 
-\sum_{\kappa''}
(\tilde{\Phi}^{\mathrm{old}}_\textrm{ax})_{\kappa\alpha,\kappa''\beta}
,~\label{eq:correction_rotational}
\end{equation}
and this correction can be back-transformed in the sought 
$\Delta^{(1)}_{\kappa\alpha,\kappa\beta}$
quantity, taking into account that the symmetric part of the first-order dynamical matrix does not need any correction.

Such procedure can be envisioned for the other
dimensionalities, and also
can be applied to treat the third line of
Eq.~(\ref{eq:variable_vector}),
but these will not be worked out in the present paper, for sake of brevity. At this stage, whatever the technique, the corrections to the vector 
Eq.~(\ref{eq:variable_vector}) needed to fulfill the constraints are known, and the second step of the procedure can be described.

From Eqs.~\ref{eq:2dcase} and~\ref{eq:torque}, the relative position of atoms multiplies the zone-center IFCs but not their derivatives; this will tend to put most of the correction on the latter quantities. To better balance the corrections between the two, we instead work with the derivatives of the IFCs in reduced coordinates; this simply adds the lattice vectors in the least-square matrix as a prefactor. 
 Once the corrections at the Brillouin zone center have been computed internally using the pseudo-inverse approach ($\Delta^{(0)}_{\kappa\alpha,\kappa'\beta}$ for the IFC corrections, $\Delta^{(1)}_{\kappa\alpha,\kappa'\beta,\gamma}$ and $\Delta^{(2)}_{\kappa\alpha,\kappa'\beta,\gamma\delta}$ for their first and second derivatives in reduced coordinates, respectively), they have to be propagated to the whole Brillouin zone taking into account the crystal symmetries. Here, we decide to put all the corrections on the $\kappa$-$\kappa'$ atomic pair within the first unit cell ($\mathbf{b}=\mathbf{0}$) and its symmetric images for $\Delta^{(0)}_{\kappa\alpha,\kappa'\beta}$:
\begin{equation}
{\tiny
\Delta_{\kappa\alpha,\kappa'\beta}^{(0)}(\mathbf{q})=   \frac{1}{N_{\mathrm{sym}}} \sum_{m} S^m_{\alpha\alpha'} S^m_{\beta\beta'}\Delta^{(0)}_{\mu\alpha',\mu'\beta'} e^{-2\pi j\mathbf{q}.(\mathbf{R}^{\mathbf{b}'}_{\mu'}-\mathbf{R}^{\mathbf{a}'}_{\mu})}} \label{DeltaC0}
\end{equation}
with $N_{\mathrm{sym}}$ the number of symmetries, $S^m_{\alpha\alpha'}$ the symmetry matrix of the crystal symmetry $m$ that projects the atom $\kappa$ ($\kappa'$) on the symmetric atom $\mu$ ($\mu'$). 
Finally, $\mathbf{a}'$ and $\mathbf{b}'$ are the cell indices of the images of the atom in cell $\mathbf{0}$ after the symmetry operation. 
To correct the IFCs derivatives, the correction is applied on the two images of the $\kappa'$ atom along the $\pm S^m_{\gamma\gamma'} \mathbf{R}^{\gamma}$ lattice vectors:
\begin{align}
\Delta _{\kappa\alpha,\kappa'\beta}^{(1)}(\mathbf{q})=&
\frac{-1}{2N_{\mathrm{sym}}} \sum_{m} S^m_{\alpha\alpha'} S^m_{\beta\beta'}\Delta^{(1)}_{\mu\alpha',\mu'\beta',\gamma} 
\nonumber
\\  
&e^{-2\pi j\mathbf{q}.(\mathbf{R}^{\mathbf{b}'}_{\mu'}-\mathbf{R}^{\mathbf{a}'}_{\mu})}
\left[e^{-2\pi j\mathbf{q}.\mathbf{R}^{m}_{\gamma}}-e^{2\pi j\mathbf{q}.\mathbf{R}^{m}_{\gamma}}\right], \label{DeltaC1}
\end{align}
with $\textbf{R}^m_\gamma=\sum_{\gamma'} S^m_{\gamma\gamma'}\mathbf{R}_{\gamma'}$.
By construction, this second term does not change the zone-center IFCs -only their derivatives- since the difference of the complex exponentials (term in brackets) resumes to a sine with a complex $2j$ prefactor. 
Finally, for the 
second-derivative corrections:
\begin{multline}
\Delta _{\kappa\alpha,\kappa'\beta}^{(2)}(\mathbf{q})= -\frac{1}{8N_{\mathrm{sym}}} \sum_{m} S^m_{\alpha\alpha'} S^m_{\beta\beta'}\Delta^{(2)}_{\mu\alpha',\mu'\beta',\gamma\delta} \\ e^{-2\pi j\mathbf{q}.(\mathbf{R}^{b'}_{\mu'}-\mathbf{R}^{a'}_{\mu})} \left[e^{2\pi j\mathbf{q}.(\mathbf{R}^m_{\gamma}+\mathbf{R}^m_{\delta})}+e^{-2\pi j\mathbf{q}.(\mathbf{R}^m_{\gamma}+\mathbf{R}^m_{\delta})}\right.\\
\left. -e^{2\pi j\mathbf{q}.(\mathbf{R}^m_{\gamma}-\mathbf{R}^m_{\delta})}-e^{-2\pi j\mathbf{q}.(\mathbf{R}^m_{\gamma}-\mathbf{R}^m_{\delta})}\right].
\end{multline}
The corrections to the second derivatives are thus propagated further in real space than the first derivatives and neither change the zone-center IFCs nor their first derivatives. 
The corrected IFCs are then
\begin{equation}
\tilde{\Phi}^{\text{new}}_{\kappa\alpha,\kappa'\beta}(\mathbf{q})= \tilde{\Phi}^{\text{old}}_{\kappa\alpha,\kappa'\beta}(\mathbf{q}) +\sum_i \Delta^{(i)}_{\kappa\alpha,\kappa'\beta}(\mathbf{q}). \label{eq:pseudoinverse2}
\end{equation}
One can check that these IFCs now properly respect both translational and rotational invariances.
Finally, we would like to mention that there are two effective ways to compute IFCs derivatives: one can estimate them based on the real-space IFCs moment,
Eq.~(\ref{eq:rotInv2}) after Fourier transform of the reciprocal IFCs, or using the long-wavelength driver of ABINIT~\cite{Royo2022}. This second approach has the advantage that it does not require to converge the IFCs derivatives with respect to the phonon wavevector mesh density for the Fourier transform of the IFCs.

\section{Long-range electrostatic of IFCs}
In general for semiconductors and insulators, the IFCs show a long-range decay in real space due to dipole-dipole interactions and higher-order long-range electrostatic interactions. 
This leads to a non-analytic behavior close to the zone center, with a direction-dependent LO-TO splitting~\cite{Gonze1997}. This creates numerical problems if the IFCs are Fourier interpolated in a brute-force manner. 
Typically, the issue is solved by splitting the total IFCs into its long-range (electrostatics) and its short-range (remaining) parts~\cite{Gonze1997}. The latter can be safely interpolated using a discrete Fourier transform, while an analytical expression is used for the long-range part.
In this section, we discuss the impact of this non-analytical behavior on the imposition of rotational invariance.

When the IFCs derivatives are estimated based on real-space moments, the treatment of the long-range electrostatics is evidently critical to get an accurate estimation of the real-space moments and then to properly impose rotational invariance. More unexpectedly, when looking at the IFCs derivatives from the long-wavelength driver, we also observe a significant breaking of the rotational invariance acting on the second derivatives of IFCs for low-dimensional materials along periodic directions,
Eq.~(\ref{eq:d2dq}), while the condition along the confined direction or on the first derivatives were initially more satisfied,
Eq.~(\ref{eq:2dcase}).
We compare in Tab.~\ref{tab:moment} the value of the second moment of IFCs to the second derivatives obtained from the long wavelength driver for phosphorene (computational details provided in Sec.~V). Increasing the convergence parameters was found to influence by a negligible manner this breaking. This suggests that some contributions are missing from the second IFCs derivatives, with the contribution from the long-range electrostatics appearing as an apparent cause. 

\begin{table}
\begin{tabular}{c|c|c}
\hline
&$ \sum_{\kappa',\beta\gamma}M^{(1)}_{\kappa\alpha,\kappa'\beta,\gamma} \epsilon_{\beta\gamma\delta}$ & $\sum_{\kappa',\beta\gamma}\tilde{\Phi}^{(1)}_{\kappa \alpha,\kappa' \beta,\gamma} \epsilon_{\beta\gamma\delta}$\\  
& [Hartree/Bohr]& [Hartree/Bohr]\\
\hline
$(\kappa\alpha,\delta)=(1x,y)$& -0.0845 &  0.0847 \\
$(\kappa\alpha,\delta)=(1x,z)$&
0.0000&0.0001 \\
$(\kappa\alpha,\delta)=(1y,x)$&
0.1180&-0.1182 \\
$(\kappa\alpha,\delta)=(1z,x)$&0.0047 & -0.0048 \\
\hline
&$- \sum_{\kappa\kappa'}M^{(2)}_{\kappa\alpha,\kappa'\beta,zz}/\Omega_0$ & $  \sum_{\kappa\kappa'}\tilde{\Phi}^{(2)}_{\kappa z,\kappa' z,\alpha\beta}/\Omega_0$\\
&[GPa] & [GPa] \\
\hline
 $(\alpha,\beta)=(x,x)$& 0.1946 & 0.0671 \\
 $(\alpha,\beta)=(y,y)$ & 0.2717 & 0.1149 \\ 
 $(\alpha,\beta)=(z,z)$ &0.7767 & 0.7767 \\
 \hline
\end{tabular}
\caption{Comparison between the IFCs derivatives and their real-space moments in the case of phosphorene, divided by the unit cell volume $\Omega_0$ to get stress units in the case of the second-order condition of rotational invariance. By Eqs.~\ref{eq:2dcase} and~\ref{eq:d2dq}, the reported quantities should be close to one another (for the second-order condition) or their sum close to 0 (for the first-order condition) to ensure rotational invariance. For the first-order condition, we only report the values of the tensor components that are higher than 10$^{-5}$ Hartree/Bohr and non-equivalent by symmetries.
}\label{tab:moment}
\end{table}


In 3D, the long-range electrostatics are approximated with a multipole expansion, where the dipole-dipole leading term is well-established following the Ewald summation rule~\cite{Gonze1997}:
\begin{multline}
\tilde{\Phi}_{\kappa\alpha,\kappa'\beta}^{\textrm{dd}}(\mathbf{q}) = \frac{4\pi}{\Omega} \sum_{\mathbf{G}\neq \mathbf{q}} \frac{\big(\mathbf{Z}^*_{\kappa\alpha}.(\mathbf{G}+\mathbf{q})\big) 
\big(\mathbf{Z}^*_{\kappa'\beta}.(\mathbf{G}+\mathbf{q})\big)}{(\mathbf{G}+\mathbf{q})\underline{\underline{\epsilon}}^{\infty}(\mathbf{G}+\mathbf{q})} \\
 e^{2\pi j(\mathbf{G}+\mathbf{q}).(\mathbf{R}_{\kappa'}-\mathbf{R}_{\kappa})} e^{-2\pi^2 \Lambda^2(\mathbf{G}+\mathbf{q})\underline{\underline{\epsilon}}^{\infty}(\mathbf{G}+\mathbf{q}) }\\ \label{eq:ewald}
\end{multline}
where $\Omega$ is the unit cell volume, $\mathbf{G}$ are linear (integer) combinations of the reciprocal lattice vectors, $Z^*_{\kappa\alpha,\gamma}$ the Born effective charge tensor, $\epsilon^{\infty}_{\gamma\delta}$ the (optical) dielectric tensor, and $\Lambda$ the Gaussian broadening intrinsic to the Ewald summation rule. 
Here, we assume that the broadening in reciprocal space is sufficiently large such that the convergent real-space contribution of the Ewald summation can be neglected. At the zone center, the IFC long-range dipole-dipole interaction is not directly translationally invariant; this condition is typically imposed on the diagonal (on-site) IFCs by~\cite{Gonze1997}:
\begin{equation}
    \tilde{\Phi}_{\kappa\alpha,\kappa'\beta}^{\textrm{dd,new}}(\mathbf{q}) = \tilde{\Phi}_{\kappa\alpha,\kappa'\beta}^{\textrm{dd}}(\mathbf{q}) - \delta_{\kappa\kappa'}\sum_\mu\tilde{\Phi}_{\kappa\alpha,\mu\beta}^{\textrm{dd}}(\mathbf{0}), \label{eq:ewald_asr}
\end{equation}
thus translational invariance is imposed on the short- and long-range parts separately~\cite{Gonze1997}. The sum over atoms of the IFCs near the Brillouin zone-center can be expressed as
\begin{equation}
 \sum_{\kappa'}\tilde{\Phi}_{\kappa\alpha,\kappa'\beta}^{\textrm{dd,new}}(\mathbf{q}\rightarrow 0)\sim\frac{4\pi}{\Omega}
\sum_{\kappa'} \frac{(\mathbf{Z}^*_{\kappa\alpha}.\mathbf{q}) 
(\mathbf{Z}^*_{\kappa'\beta}.\mathbf{q})}{\mathbf{q}\underline{\underline{\epsilon}}\mathbf{q}},
\end{equation}
taking into account that $|\mathbf{q}|<<|\mathbf{G}|$. 

\begin{figure}[htp!]
\includegraphics[width=0.48\textwidth]{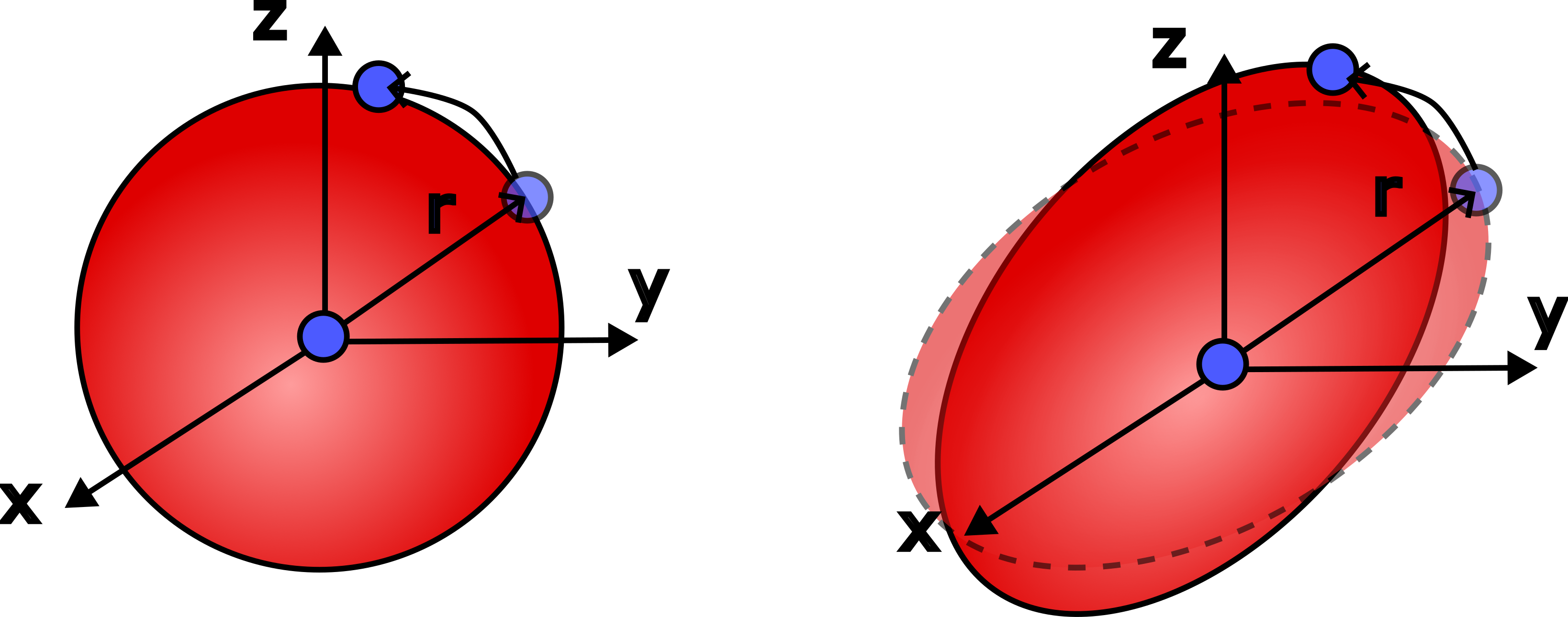}
\caption{Schematic representation of potential isosurface (red) due to the electrostatic interactions between point charges in the case of (left) isotropic dielectric environment (right) anisotropic dielectric environment. The energy is trivially invariant under a global rotation of the system in the isotropic case, while the second case requires also the rotation of the dielectric tensor.}
\label{fig:pointcharge}
\end{figure}

Let us put the Ewald summation approach in perspective with rotational invariance and let us examine more precisely the role of the dielectric tensor. To do so, we consider the simpler case of the electrostatics interacting of point charges in an anisotropic environment; their IFCs are
\begin{multline}
\Phi_{\kappa\alpha\kappa'\beta} = \sum_{AB}\frac{Z_{A} Z_{B}}{2 M D^3_{AB}} \left(3\frac{\Delta_{AB\alpha}\Delta_{AB\beta}}{D^2_{AB}} - \epsilon^{-1}_{\alpha\beta}\right) \\
(\delta_{\kappa A}-\delta_{\kappa B})(\delta_{\kappa' A}-\delta_{\kappa' B}),\label{eq:ifc_born}
\end{multline}
 where $Z_A$ is the charge of atom A, $\boldsymbol{\Delta}_{\kappa\kappa'} = \underline{\underline{\epsilon}}^{-1} \Delta \mathbf{R}_{\kappa\kappa'}$, $D_{\kappa\kappa'} = \Delta\mathbf{R}_{\kappa\kappa'}. \boldsymbol{\Delta}_{\kappa\kappa'}$ and $M=(\det(\underline{\underline{\epsilon}}))^{0.5}$. Upon a global rotation of the system around the atom $\kappa$, there is a non-zero resulting force, due to the anisotropy of the dielectric tensor
 (except if the latter is rotated at the same time). In consequence, the system is not invariant under rotation. This effect is illustrated in Fig.~\ref{fig:pointcharge}, where the effect of a rotation of electrostatic potential isosurface in an anisotropic dielectric medium is shown graphically. 

These findings must hold when considering a 3D periodic crystal with an anisotropic dielectric tensor. 
Note that this is a consequence and a limitation of the electrostatic model, where the invariance is lost with the introduction of Born effective charges and dielectric tensors which do not relay any dependence on the atomic positions. 
Physically, when rotating the atoms of a real crystal, the electronic cloud rotates at the same time and thus the corresponding dielectric tensor and Born effective charges accordingly. 
Such a model would lead to a vanishing contribution to the rotational invariance breaking, which would be consistent with the observation that the breaking of rotational invariance is small for its first-order condition without including a contribution from electrostatics.
Therefore, using in this case the first derivatives from the long-wavelength driver would be exact. 
The statement would however not hold for the second derivatives. 
Extending the electrostatic model to account for this effect is not trivial and left for future work. 
In addition to those considerations for the dipole-dipole interactions, higher-order electrostatics terms (dipole-quadrupole, quadrupole-quadrupole, ...) also contribute to the long-range decay of IFCs~\cite{Stengel2013,Royo2019,Royo2022}.
They have already been shown to impact phonon dispersion close to the Brillouin zone center~\cite{Brunin2020,Royo2020b}. 
With the present electrostatic model, we face however the same problem for rotational invariance as for the dipole-dipole case due to the anisotropy of the dielectric tensor.

To the best of our understanding, the most meaningful way to compute a phonon band structure is to first impose rotational invariance including the IFCs first derivatives from the long-wavelength driver. 
Then, use the standard procedure for the interpolation with the splitting of the IFCs into their short- and long-range parts~\cite{Gonze1997}. 
Because of the large breaking of the second-order condition, Eq.~(\ref{eq:d2dq}) and the absence of rotational invariance for the long-range electrostatics, we believe it does not yet make sense to impose the corresponding conditions on the zone-center IFCs or their second derivatives currently.
In practice, we impose first-order condition of rotational invariance based on the coarse-mesh computed IFCs following 
Eq.~(\ref{eq:pseudoinverse}) as well as Eqs.~(\ref{DeltaC0}) to~(\ref{eq:pseudoinverse2}), then separate these corrected IFCs into their short- and long-range parts:
\begin{equation}
\tilde{\Phi}^{\text{new,sr}}_{\kappa\alpha,\kappa'\beta}(\mathbf{q}_\mathrm{c})= \tilde{\Phi}^{\text{new}}_{\kappa\alpha,\kappa'\beta}(\mathbf{q}_\mathrm{c})-\tilde{\Phi}^{\text{dd,new}}_{\kappa\alpha,\kappa'\beta}(\mathbf{q}_\mathrm{c})
\end{equation}
with the long-range part given by Eqs.~(\ref{eq:ewald}) and~(\ref{eq:ewald_asr}), and $\mathbf{q}_\mathrm{c}$ belonging to the coarse phonon mesh. Then, the short-range IFCs are Fourier interpolated and the full IFCs on the dense phonon mesh $\mathbf{q}_\mathrm{d}$ are obtained following
\begin{equation}
\tilde{\Phi}^{\text{new}}_{\kappa\alpha,\kappa'\beta}(\mathbf{q}_\mathrm{d})= \tilde{\Phi}^{\text{new,sr}}_{\kappa\alpha,\kappa'\beta}(\mathbf{q}_\mathrm{d})+\tilde{\Phi}^{\text{dd,new}}_{\kappa\alpha,\kappa'\beta}(\mathbf{q}_\mathrm{d}).
\end{equation}

Finally, we would like to discuss the role of long-range electrostatics to describe low-dimensional materials. 
For 2D and 1D systems, several correction schemes have been proposed ~\cite{Sohier2016, Sohier2017,Royo2021,Rivano2023,Rivano2024} which resemble the Ewald summation approach previously discussed. Those scheme conclude to the absence of LO-TO splitting and a linear dependence with respect to $q$ near the Brillouin zone center. These formalisms yield no contribution to the IFCs derivatives. One could then simply assume that the treatment of the long-range electrostatics does not matter in this case in contrast to 3D. 
However, we argue here the contrary, since the long-range electrostatics term should only be invariant under rotation if the dielectric tensor and the Born effective charges are rotated at the same time. 
This suggests the corresponding models should be refined to include this consideration, similarly to the 3D case.

\section{Validation}

\begin{figure}[htp!]
\includegraphics[width=0.48\textwidth]{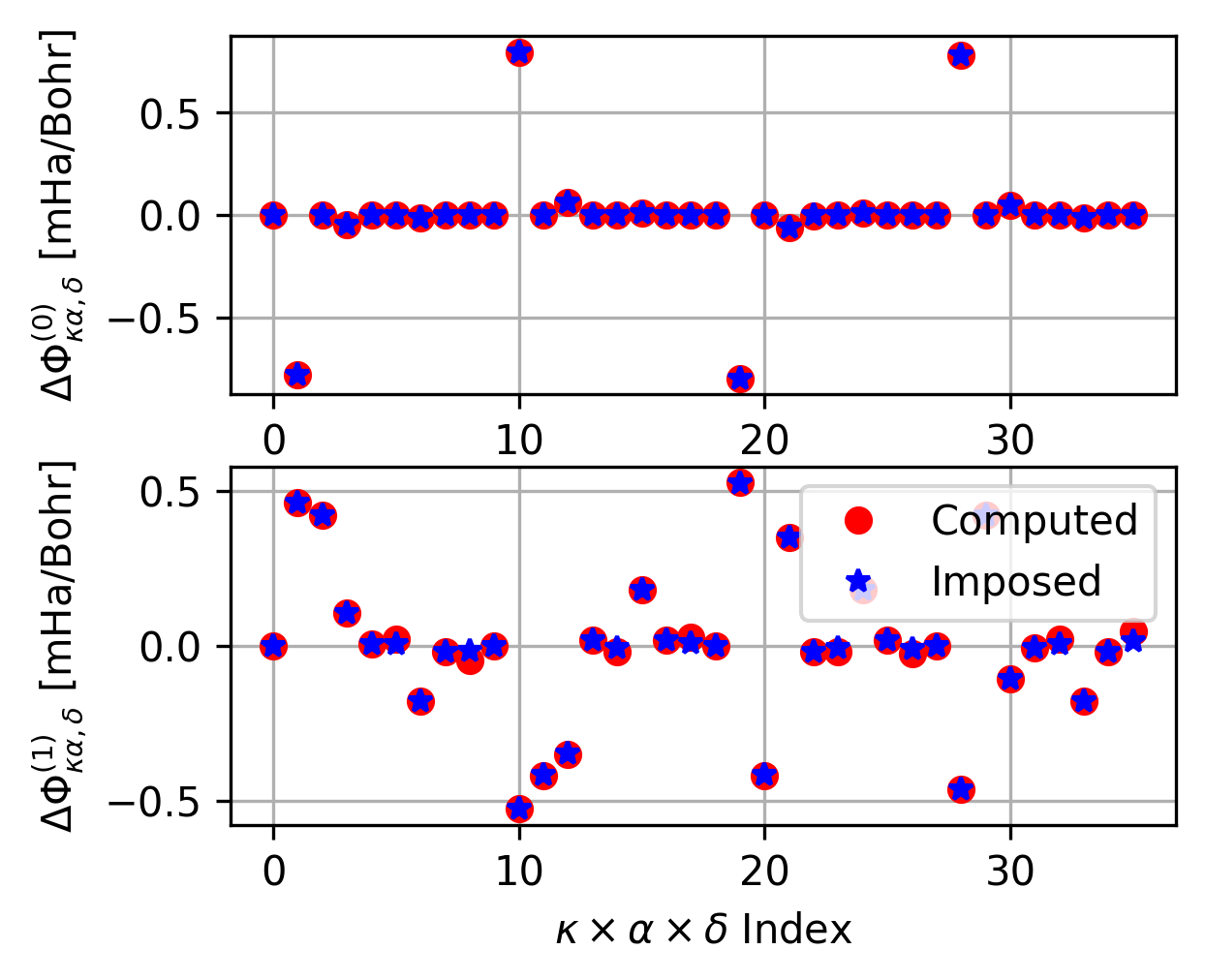}
\caption{Comparison for phosphorene of (top) the correction to the zone-center torque 
$\Delta\Phi^{(0)}_{\kappa\alpha,\delta}$
arising from the zone-center corrections $\Delta^{(0)}_{\kappa\alpha\kappa'\beta}$
from the pseudoinverse (Eq.~(\ref{eq:phi0}), red circle), or after symmetrization of the correction, (Eq.~(\ref{DeltaC0}) then taking $\mathbf{q}=0$, Eq.~(\ref{eq:phi0}), blue stars). (Bottom) same but now for the derivative corrections, $\Delta^{(1)}_{\kappa\alpha\kappa'\beta}$, following Eqs.~(\ref{eq:phi1}) and~(\ref{DeltaC1}) (same symbol convention).
}
\label{fig:compdc}
\end{figure}

 To validate the present implementation, we first compute the correction to the zone-center torque and to the IFC derivatives from the Moore-Penrose pseudoinverse output ($\Delta^{(0)}_{\mu\alpha',\mu'\beta'}$ and $\Delta^{(1)}_{\mu\alpha',\mu'\beta',\gamma}$) in the case of phosphorene. As mentioned in the previous section, without a proper treatment of rotational invariance for the long-range electrostatics, the breaking of rotational invariance on the second derivatives of the IFCs is too large to be physical, and therefore we simply consider the corrections on the first derivatives of the IFCs hereafter.
 The IFCs derivatives are estimated here from the real-space IFCs moment. These corrections contribute to the rotational invariance conditions following:
 \begin{equation}
 \Delta\Phi^{(0)}_{\kappa\alpha,\delta}=\sum_{\kappa',\beta\gamma} \Delta^{(0)}_{\kappa\alpha,\kappa'\beta}(R^0_{\kappa'\gamma}-R^0_{\kappa\gamma})\epsilon_{\beta\gamma\delta}, \label{eq:phi0}
 \end{equation}
 and
  \begin{equation}
\Delta\Phi^{(1)}_{\kappa\alpha,\delta}=\sum_{\kappa',\beta\gamma} R_{\gamma'\gamma}\Delta^{(1)}_{\kappa\alpha,\kappa'\beta,\gamma'}\epsilon_{\beta\gamma\delta}, \label{eq:phi1}
 \end{equation}
 with $R_{\gamma'\gamma}$ the lattice vectors converting the q-component of the tensor from reduced to cartesian coordinates.
 Then, we compare these torques to the ones obtained after the IFCs imposition following the expressions Eqs.~(\ref{DeltaC0}) and~(\ref{DeltaC1}) with the same expression taking $\mathbf{q}=0$. The results are shown in Fig.~\ref{fig:compdc}. We find a nearly perfect agreement for most indices, with only small discrepancies for very small IFCs corrections. 

\begin{figure}[htp!]
\includegraphics[width=0.38\textwidth]{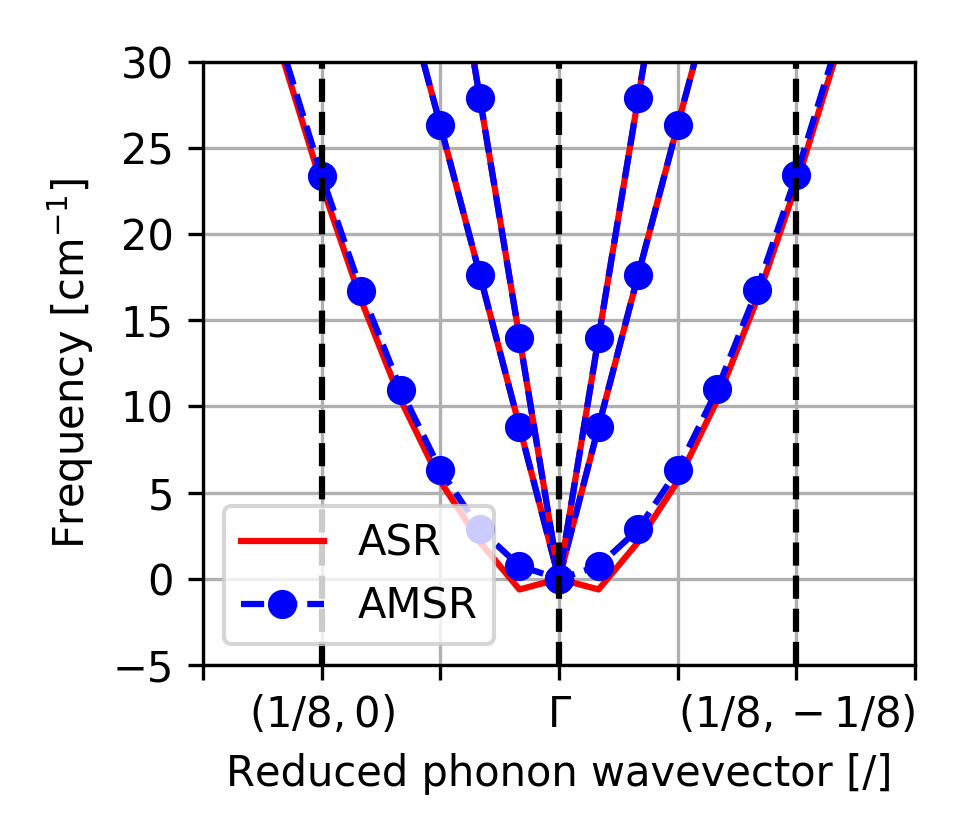}
\caption{Validation of symmetries preservation upon application of the rotational invariance in the case of monolayer WS$_2$ for a 6$\times$6 computed phonon wavevector grid. We compare the acoustic sum rule (ASR, full red line) to the case where both translational and rotational invariances are imposed at the same time (AMSR, dash dotted blue line). Along two equivalent high-symmetry lines (ticks shown in reduced coordinates) the phonon frequencies are equivalent. }
\label{fig:ws2_sym}
\end{figure}

 We have also checked that the added correction was correctly respecting the crystal symmetries, as shown in Fig.~\ref{fig:ws2_sym} for monolayer hexagonal WS$_2$. Indeed, following two equivalent high-symmetry lines, we recover the same phonon frequencies, providing a check that the symmetries are properly implemented. Note that for WS$_2$ with ONCVPSP pseudopotentials~\cite{pseudodojo2018}, the breaking of the first-order condition of rotational invariance is small, and consequently, we will not discuss it further down in the manuscript.

To understand and illustrate the impact of rotational invariance on the IFCs, we first consider the situation where only the zone-center IFCs are modified to impose rotational invariance. With the introduction of
\begin{equation}
    b_{\kappa\alpha,\delta}=\sum_{\kappa',\beta\gamma}\Im(\Phi^{(1)}_{\kappa\alpha,\kappa'\beta,\gamma}) \epsilon_{\beta\gamma \delta}
\end{equation}
that we express into its vectorized form, the corrections to be applied on the IFCs can then be written as
\begin{equation}
\mathbf{\Delta} =   \xi\underline{\underline{A}}^+ \mathbf{b}- \underline{\underline{A}}^+ \underline{\underline{A}} \mathbf{\tilde{\Phi}} \label{eq:pseudo}
\end{equation}
where $\mathbf{\tilde{\Phi}}$ is the zone-center IFCs expressed in their vectorized form. The condition matrix $\underline{\underline{A}}$ is adapted to account that only the zone-center IFCs are taken here as variable for the pseudo-inverse. Finally, $\xi$ is a parameter that we vary; when $\xi=1$, the first IFCs derivatives are used in their entirety when imposing rotational invariance. When smaller than unity, only a fraction of these derivatives are used upon the imposition, meaning the zone-center IFCs must be further corrected to impose rotational invariance. 

\begin{figure}[htp!]
\includegraphics[width=0.48\textwidth]{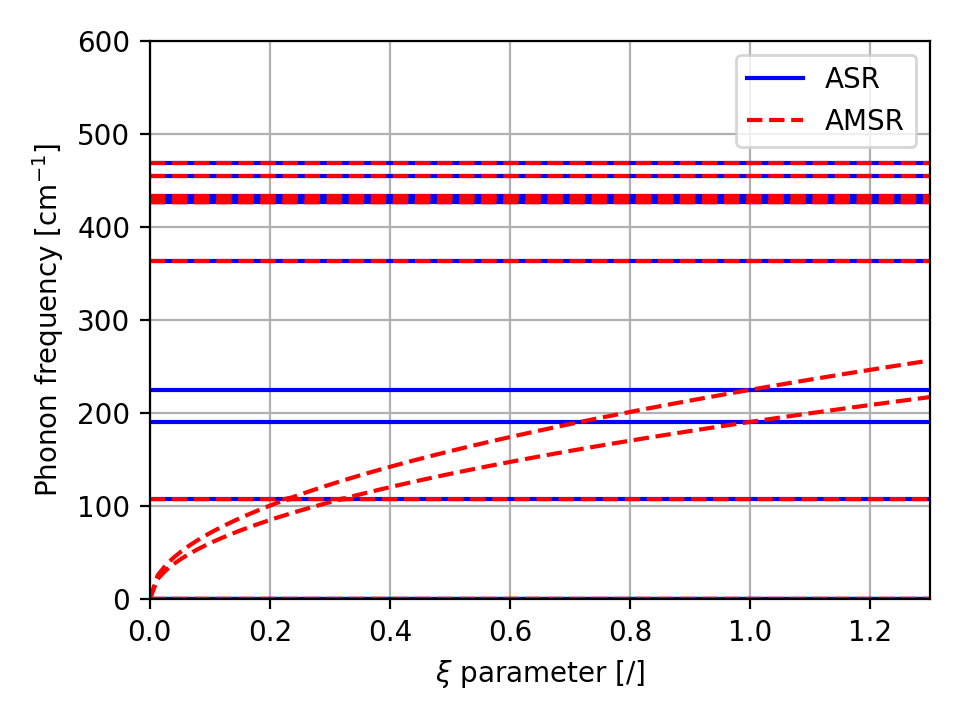}
\caption{Phonon frequencies in phosphorene considering either (dash blue) only translational invariance or both translational and rotational invariance (red) with varying fraction of the IFCs derivatives included when the rotational invariance is imposed 
(Eq.~(\ref{eq:2dcase})).}
\label{fig:phospho1}
\end{figure}

We illustrate 
as an example the impact of rotational invariance on the zone-center phonons for phosphorene in Fig.~\ref{fig:phospho1}, where we vary the $\xi$ parameter (computational method presented in the next section)
and impose the rotational invariance on the zone-center IFCs accordingly. 
As one can see, two specific modes (corresponding to the out-of-plane optical mode with eigendisplacement components along either the armchair or zigzag directions) are impacted, and go to zero when one simply removes the contribution corresponding to the IFCs derivatives, for which we somewhat recover the solution for a non-periodic case, i.e. a molecule. 
One can see that for the examined example, with IFCs originating from DFT calculations, the imposition of rotational invariance only requires a small change of the IFCs at the zonecenter, or of the IFCs derivatives.

\section{Computational details and application}

All computations are performed with the ABINIT software package~\cite{Abinit2005,Abinit2009,Gonze2016,Romero2020}. 
We started from ABINIT version 10.5.6, https://github.com/abinit/abinit/releases/tag/10.5.6 , then implemented Eq.~(\ref{eq:pseudoinverse}) as well as Eqs.~(\ref{DeltaC0}) to~(\ref{eq:pseudoinverse2}).
These modifications will be available in ABINIT version 10.8.X.

We use the local-density approximation (LDA) with either Optimized Norm-Conserving Vanderbilt Pseudopotentials (ONCVPSP, v0.4.1)~\cite{Hamann2013,pseudodojo2018} or Hartwigsen Goedecker Hutter (HGH)~\cite{Hartwigsen1998} pseudopotentials. The formers have non-linear core corrections, preventing currently the computations of the IFCs derivatives with the long-wavelength driver of ABINIT, while the latters do not, allowing such computations. 
However, the HGH pseudopotential requires a higher plane-wave cut-off energy to reach convergence. 
In terms of the $\Delta$ factor~\cite{Lejaeghere2016}, HGH pseudopotentials perform relatively well, with $\Delta < 2$ meV/atom, for most of the elements considered in this work (H, C, P, O). 

The Brillouin zone is sampled with a zone-centered mesh with 16 points along each periodic direction and a single point in the confined directions. The structures of the considered materials (lattice parameters and internal degrees of freedom) are first optimized with a Broyden–Fletcher–Goldfarb–Shanno algorithm until the forces on the atoms are smaller than $10^{-5}$ Hartree/Bohr. The IFCs and subsequent phonon band structure are obtained with the Density Functional Perturbation Theory (DFPT) formalism~\cite{Gonze1997}.
The IFCs derivatives are either estimated based on the real-space IFCs moment, or obtained as an output by the long-wavelength driver of ABINIT~\cite{Royo2022}. Since the breaking of translational and rotational invariances typically originates from the estimation of the exchange-correlation functional on a finite grid in real space, linked to the plane-wave basis set, and from the non-linear core corrections, we discuss later the convergence of the phonon band structures with respect to the plane-wave cut-off energy. 
We also do not consider any Coulomb truncation of the DFT potentials typically used to avoid spurious interactions from periodic images in the out-of-plane direction~\cite{Sohier2016}. 
To minimize the error made by this approximation, we specifically study materials which should not yield any problematic behaviors, either because their Born charges are nearly negligible (polyethylene chain, 1D case) or because the flexural mode is not impacted by the long-range electrostatics, due to the effective charge symmetries (phosphorene, 2D case). 
As an additional check, the proper convergence with respect to the phonon wavevector mesh sampling is investigated. For the long-range electrostatics in the 2D case, we use our recent implementation in ABINIT~\cite{VanTroeye2025} for the dipole-dipole interactions. Following the discussion of Sec.~III, we do not impose the second-order condition of rotational invariance since we believe it requires a rotationally-invariant expression for the long-range electrostatics.

We tested the implementation on different system dimensionalities, starting with H$_2$O and CO molecules for the 0D case. We used ONCVPSP pseudopotentials with a 55~Ha cut-off energy and embedded the molecules in a cubic box of 20~Bohr-lattice parameter. The results are listed in Tab.~\ref{tab:mol}. As shown, the breaking can be substantial ($\sim$5-30~cm$^{-1}$) when only the translational invariance is enforced. The implementation of rotational invariance properly imposes that 2 or 3 additional modes vanish depending on the molecule linearity.

\begin{table}[h]
\begin{tabular}
{ r |c| c  c  c  c  c  c }
  Molecule & Invariance & \multicolumn{6}{c}{Vibrational frequency [cm$^{-1}$]} \\
  \hline
 \multirow{ 2}{*}{H$_2$O}&ASR  & -29.3 & -24.3 & 16.0 & 1601 & 3699 & 3811\\
 &AMSR & 0& 0 & 0 &  1599& 3698&3812 \\
  \hline
 \multirow{ 2}{*}{CO} &ASR & -5.97 & -5.97 & 2171 \\
 & AMSR & 0 & 0 & 2171 \\
 \hline
\end{tabular}
\caption{\label{tab:mol}Vibrational frequencies of H$_2$O and CO molecules with only translational invariance imposed (ASR) or with rotational invariance as well (AMSR). The translational modes (that are close to 0 with ASR, exactly 0 with AMSR) are omitted for sake of conciseness.}
\end{table}

\begin{figure}[htp!]
\includegraphics[width=0.43\textwidth]{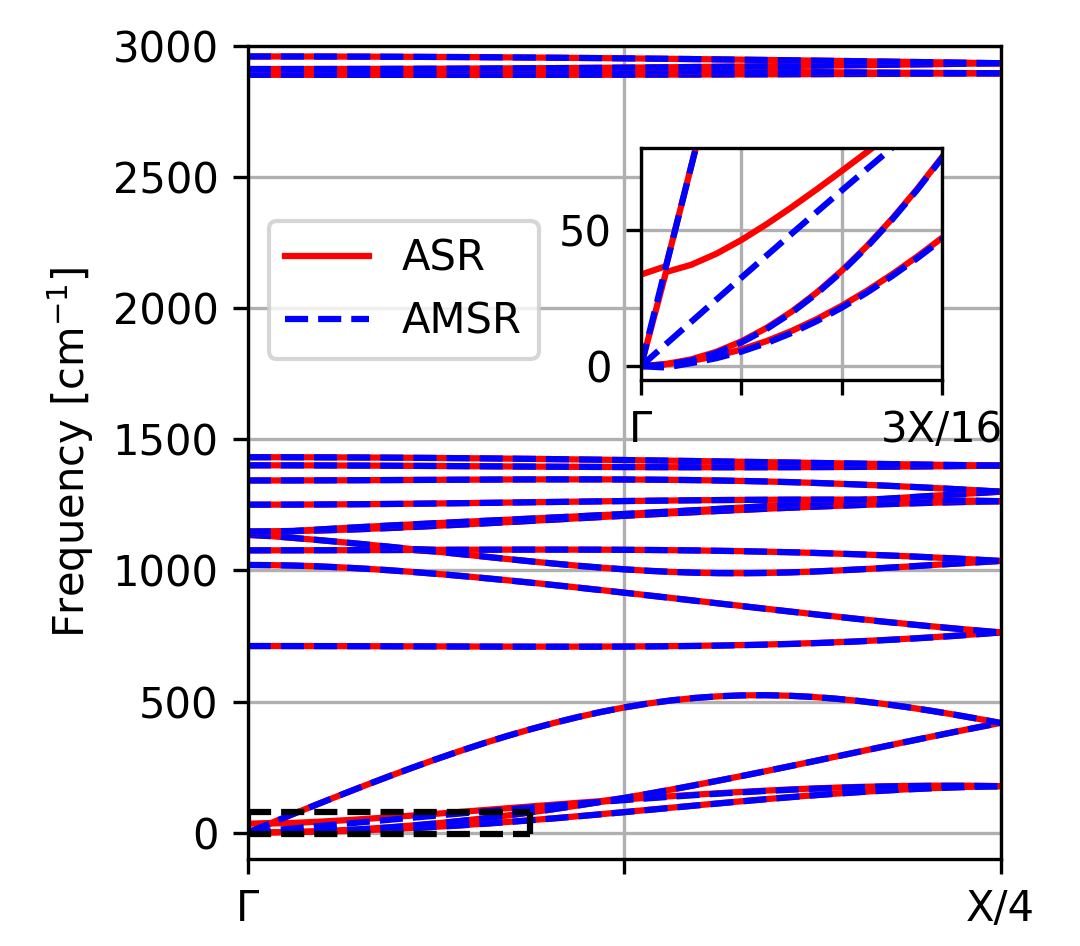}
\caption{Phonon modes of a polyethylene chain with translational invariance (ASR, full red) or translational + rotational invariance (AMSR, dash blue). The inset shows the acoustic and rotational phonon modes close to the Brillouin zone center (area determined by the black dashed lines).}
\label{fig:1dsystem}
\end{figure}

Next, we look to the case of a polyethylene chain as a prototype one-dimensional system. HGH pseudopotentials are used, in order to have access to the IFCs derivatives through the long-wavelength driver. 
We use 20-Bohr lattice parameters for the confined directions. 
The relaxed lattice parameter is 2.52~\AA{} along the periodic direction. 
The phonon band structure of polyethylene is converged within a few cm$^{-1}$ with a 150~Ha plane-wave kinetic cut-off  energy (see Fig.~\ref{fig:1D_qptconv} in App.~D for the corresponding convergence study). 
It is important to note that long-range electrostatics is not treated separately in this case. 
The contribution from dipole–dipole interactions is expected to be minimal, as the Born effective charges are relatively small ($<0.2e$). 
In contrast, quadrupole–dipole interactions may not be negligible ($<3e$-Bohr); however, their contribution to the IFC derivatives in one-dimensional systems remains uncertain, as only a dipole–dipole formalism has been proposed to date~\cite{Rivano2023}. 
It turns out that the convergence of the phonon band structure with respect to the q-point sampling appears smooth (see Fig.~\ref{fig:1D_qptconv}), suggesting that a detailed treatment of long-range electrostatics may not be critical in this case. We show in Fig.~\ref{fig:1dsystem} the phonon band structure when we impose rotational invariance in addition to translational invariance. 
In general, there should be four modes at the Brillouin-zone center that cancel out in 1D systems (3 translation modes + 1 rotation of the chain around its center of mass). 
One of these is incorrectly described without imposition of rotational invariance ($\sim 35$ cm$^{-1}$). 
When rotational invariance is imposed, the expected behavior is properly recovered, with the associated mode typically showing a linear dispersion close to the zone center. 
Concerning the flexural modes of the chain, they are relatively weakly impacted by the imposition of rotational invariance. 

\begin{figure}[htp!]
\includegraphics[width=0.48\textwidth]{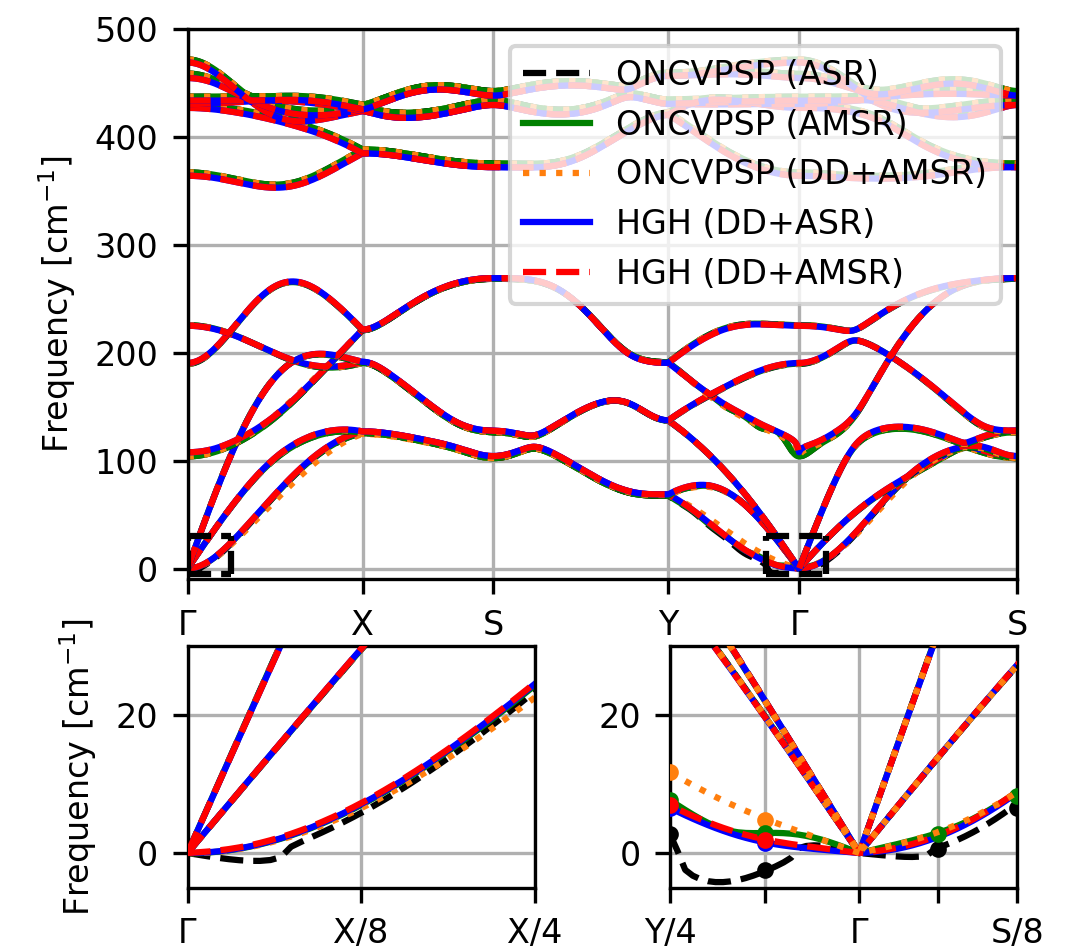}
\caption{Phonon band structure of phosphorene with translational invariance (ASR, dashed black and full blue lines) plus rotational invariance (AMSR, full green, dotted orange and red dashed lines) and for ONCVPSP or HGH pseudopotentials. Dipole-dipole (DD) corrections are added for the specified cases.}
\label{fig:2dsystem}
\end{figure}

We then move to 2D materials, with phosphorene as our main test case. We use a dielectric thickness of 5.32~\AA and consider vacuum as the embedding dielectric. The Born effective charges of phosphorene show off-diagonal components~\cite{Ponce2023}, namely $Z^*_{\kappa z,y}$ and $Z^*_{\kappa y,z}$. We thus expect here the in-plane polarization induced by the out-of-plane atomic displacements to influence the flexural mode. This material crystallizes in an orthorhombic puckered structure~\cite{Ankit2015} and both ONCVPSP and HGH pseudopotentials give comparable lattice parameters 
(3.26~\AA{} and 4.36~\AA{} for both cases).
While the flexural optical modes converge smoothly with respect to the phonon wavevector mesh sampling when dipole-dipole interactions are included, this is not entirely the case for the flexural acoustic mode. 
Including dipole-quadrupole and quadrupole-quadrupole interactions may solve the problem like it does for 3D materials ~\cite{Royo2020b}.
Importantly, we note that the flexural acoustic phonon mode exactly computed from DFT in the immediate vicinity of the zone center shows a negative value for the ONCVPSP pseudopotentials (see Fig.~\ref{fig:2dsystem}). This effect cannot be attributed to the treatment of the electrostatics, since the latter only modifies the way IFCs are interpolated. Similarly, pushing the convergence and calculation parameters (energy cut-off, k-point sampling, self-consistent cycle threshold) did not solve the issue.  

For the HGH pseudopotential with the IFCs derivatives obtained from the long-wavelength driver, the breaking of rotational invariance is found to be small and to lead to a slight increase of the group velocity in the immediate vicinity of the zone-center. 
This suggests that here, one would need to also impose the second-order condition of rotational invariance, Eq.~(\ref{eq:d2dq}), to recover a purely quadratic dispersion for the flexural acoustic mode in this case. 
Meanwhile, the imposition of rotational invariance for the ONCVPSP pseudopotentials with the real-space IFCs moments  curates the observed negative modes when the long-range IFCs electrostatics is not treated separately. In contrast, when it is treated separately, the negative mode remains. 
This can be understood from the fact that the correction schemes are not rotationally invariant by construction, as discussed previously in this manuscript.

\begin{figure}[htp!]
\includegraphics[width=0.41\textwidth]{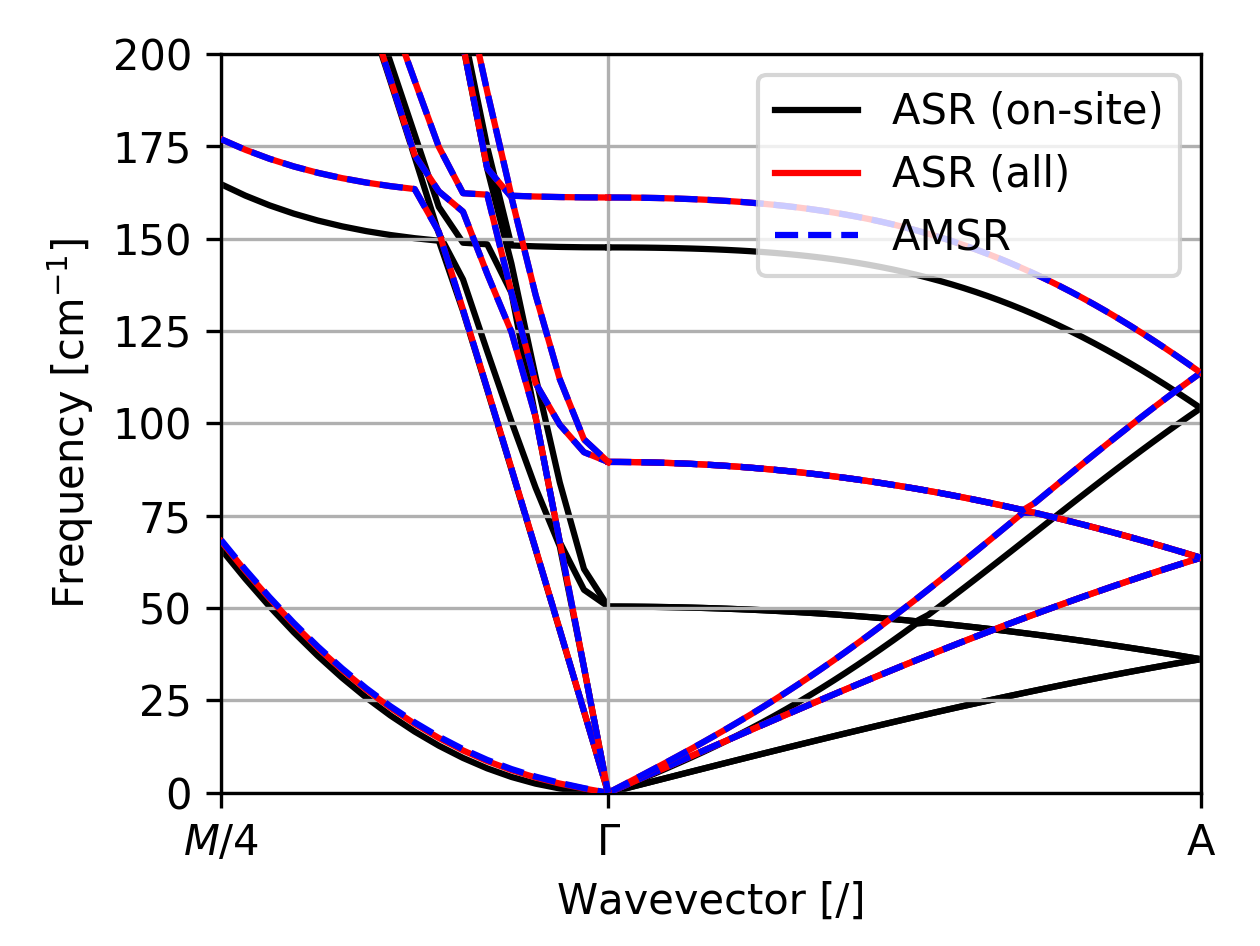}
\caption{Acoustic and low-frequency optical modes for AB-stacked graphite for the (in-plane) $\Gamma\rightarrow M$ and (out-of-plane) $\Gamma\rightarrow A$ high-symmetry lines. We show in solid black and red the phonon band structures obtained with translational invariance only, but with the correction imposed only on the on-site (black) or split between the zone-center IFCs (red). The blue-dashed curve corresponds to the case where rotational invariance is also imposed.}
\label{fig:graphite}
\end{figure}

For 3D materials, we look at graphite in the AB stacking configuration. We use the same phonon data than in Ref.~\onlinecite{VanTroeye2016} (same computational method), with GGA-PBE as the exchange-correlation functional corrected by DFT-D2 Grimme's dispersion~\cite{Grimme2006} to bind the layers together. 
The properties of graphite are strongly anisotropic, and thus expected to be an interesting test case for rotational invariance. The corresponding results are shown in Fig.~\ref{fig:graphite}.
As one can see, not only the phonon dispersion are impacted, but also the modes at the Brillouin zone center compared to the case where only the acoustic sum rule is imposed. 
We trace down the origin of this variation to the way translational invariance is imposed: indeed, the correction is traditionally imposed only on the diagonal (on-site) IFCs, see Eq.~(\ref{eq:ewald_asr}), while with the present scheme the corrections are distributed among all IFCs. 
This highlights that specific care need to be taken on the way translational invariance is imposed when considering weakly-bound materials. 
At the current stage, we cannot assess the
advantage of one with respect to the other. 
Beyond this effect, the imposition of rotational invariance has a negligible impact on the phonon dispersion in this case.

\section{Conclusion}

In this work, we have first shown how to impose the rotational invariance on first-principles dynamical matrices and interatomic force constants, through a reciprocal-space formulation, at variance with the previously proposed real-space
approach. 
Such a formulation might be more efficient in the case of a large number of atoms present in the primitive cell, as the scaling is more favourable.

Then, we revisited the impact of rotational invariance imposition on phonon band structures and examined how its influence varies with the system dimensionality. 
While its consequence on molecules is relatively straightforward - the frequency of two or three modes must vanish due to rotational invariance-, its effect in one- and two-dimensional materials is more complex, impacting both the zone-center IFCs as well as their first and second derivatives at the zone center. 
This leads to a quadratic dispersion of the flexural acoustic modes in the immediate vicinity of the Brillouin zone center. 
Some challenges remain when dealing with the long-range electrostatics IFCs, with the usual dipole-dipole interaction term not being rotationally invariant. 
In consequence, we currently only enforce the first-order condition of rotational invariance on the zone-center IFCs and their first derivatives using a Moore-Penrose pseudoinverse.

Following our implementation in the ABINIT software, we test it on several systems. While the enforcement of the rotational invariance appears to a significant impact on the zone-center rotational mode in zero- and one-dimensional systems, its effect on the flexural branches appears more scattered: minor in a polytethylene chain or phosphorene with HGH pseudopotentials, significant for phosphorene with ONCVSP pseudopotentials or graphite. The way that the translational invariance is imposed also seems to have a significant impact. 
Those observations should be further evaluated across a broader set of low-dimensional or weakly-bound materials, particularly in cases where long-range electrostatic interactions are included. In any case, seemingly minor changes may be sufficient to induce significant variations in properties derived from phonon behavior, such as electrical or thermal conductivity, where a quadratic phonon dispersion induces a divergence of the phonon population close to the zone center. 

Another promising direction would be to leverage the knowledge of IFC derivatives to accelerate convergence with respect to the q-point mesh. In low-dimensional materials, the need for dense q-point sampling is often driven by the slow convergence of flexural modes near the zone center. In principle, this requirement could be alleviated by incorporating IFC derivatives computed from the long-wavelength driver. Deriving a rotationally-invariant expression for the long-range electrostatics part of the IFCs may also be useful for polar materials.

\begin{acknowledgments}
{The authors acknowledge the Imec Industrial Affiliation Program (IIAP) for funding. This work is an outcome of the Shapeable 2D magnetoelectronics by design project (SHAPEme, EOS Project No. 560400077525) that has received funding from the FWO and FRS-FNRS under the Belgian Excellence of Science (EOS) program. This work was supported by the Fonds de la Recherche Scientifique - FNRS under Grant number J.0091.26 (CDR).  }
\end{acknowledgments}

\section{Appendix}

\subsection{Challenges with real space imposition of rotational invariance}

In this Appendix we discuss the challenges related to the imposition of rotational invariance in real space, which can be done using Eq.~(\ref{eq:rotInv}). As a reminder, in DFPT, the IFCs are computed first in reciprocal space based on a finite mesh of phonon wavevectors. The real space IFCs are then obtained using a (discrete) Fourier transform:

\begin{equation}
C_{\kappa\alpha,\kappa'\beta}(0,\mathbf{b}) \approx \frac{1}{(2\pi)^3}\sum_\mathbf{q} \tilde{C}_{\kappa\alpha,\kappa'\beta}(\mathbf{q}) e^{-2\pi j\mathbf{q}.(\mathbf{R}^b_{\kappa'}-\mathbf{R}^0_{\kappa})}.\label{eq:model}
\end{equation}

With a finite phonon wavevector mesh, we have only access to real-space IFCs up to a certain distance, in terms of $\mathbf{R^b}$, determined by the grid density along the reciprocal vectors. We can then in principle impose rotational invariance on those IFCs based on Eq.~(\ref{eq:rotInv}).

\begin{figure}[htp!]
\includegraphics[width=0.48\textwidth]{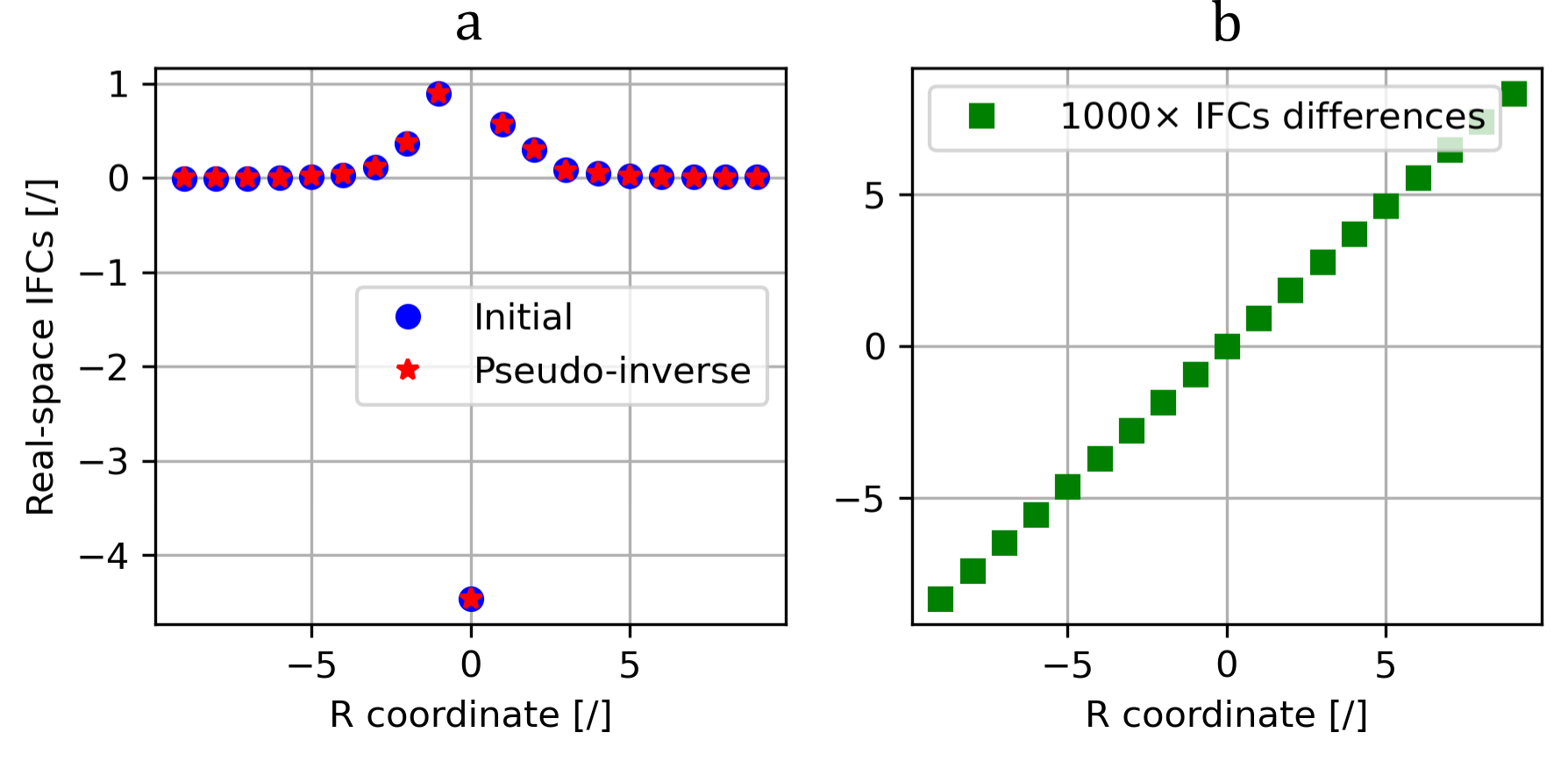}
\includegraphics[width=0.48\textwidth]{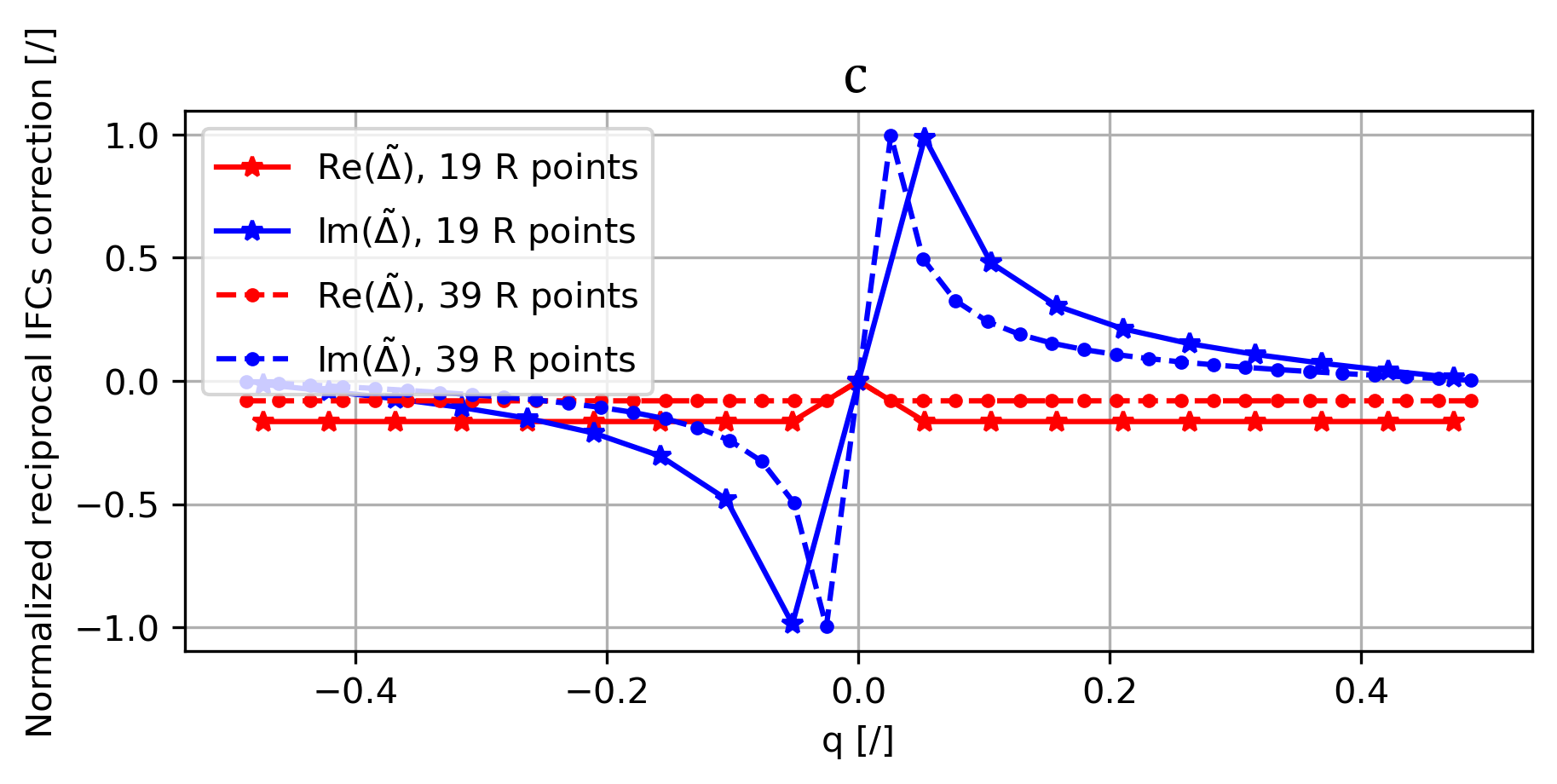}
\caption{Illustration of the challenges associated with the imposition of rotational invariance in real space with a simplified model. (a): initial IFCs with the proposed model in Eq.~(\ref{eq:model})  (blue disks) and after the imposition of conditions on the real-space IFCs (red stars). (b): corresponding corrections on the IFCs. The pseudo-inverse leads to a linear increase in the long-range IFCs in order to impose rotational invariance. (Bottom): Fourier transform of the correction for two initial sampling in real space (more points indicates that further IFCs are included). Due to the linear correction with distance in real space, the function is non-analytical close to the zone center.}
\label{fig:AMSR_real}
\end{figure}

Let us consider that we use a least-square or a pseudo-inverse to do so. In such an approach, the weight associated with the long-range IFCs will scale as $R^\mathbf{b}_{\kappa'\beta}-R^\mathbf{0}_{\kappa\beta}$, meaning that the long-range contribution will be corrected more significantly by the corresponding method. 
This is not what we want, since the physical trends for the IFCs should be a monotonous decrease at long distance. In reciprocal space, such a linear correction will then translate into the derivative of the Kronecker delta $\delta^{(1)}(\mathbf{q})$, which is not analytical.  
In addition, with increasing mesh density, the correction will be pushed further and further in real space, since further and further IFCs become available in the pseudoinverse. To illustrate these effects, we construct a simple 1D mathematical model for the real-space IFCs, based on an exponential function:
\begin{equation}
\Phi(R)=e^{-R}(1+2U(R))
\end{equation}
with $U(R)$ an uniform random distribution and $R$ the atomic position.
We then impose translational invariance (corresponding IFCs represented as blue dots in Fig.~\ref{fig:AMSR_real}a), in addition to the following condition enforced with the pseudoinverse:
\begin{equation}
\sum_R R\,\Phi(R)=0.
\end{equation}
Without any uniform random distribution, this condition is trivially imposed by the shape of the IFCs. Its shows the same long-range behavior than rotational invariance. We then define the difference between corrected $\Phi^{\text{corr}}$ and initial $\Phi^{\text{init}}$ IFCs as \begin{equation}
\Delta=\Phi^{\text{corr}}-\Phi^{\text{init}}.
\end{equation}
When analyzing the corresponding corrections, one can observe they are  preferentially put on the furthest IFCs (see Fig~\ref{fig:AMSR_real}b). The corresponding reciprocal-space corrections then show a non-analytical behavior close to the zone-center, that is amplified if one increase the number of $R$ coordinates in real-space (see Fig~\ref{fig:AMSR_real}c). 

For those reasons, we advocate for the imposition of the correction on the first and second neighbors IFCs -or on any IFCs albeit weighted with a decaying function of the distance-. This could in principle directly be implemented on top of the pseudo-inverse with the introduction of a weight matrix for the real-space IFCs. However, as discussed in the body of the manuscript, the imposition of rotational invariance on the long-range IFCs, related to electrostatics, is still challenging. Working in reciprocal space with the IFCs derivatives from the long-wavelength driver remain the safest approach to the best of our knowledge. 

\subsection{Perturbation expansion}

To understand the impact of the rotational invariance on the dispersion of the phonon modes close to the Brillouin zone center, we can use perturbation theory on the eigenvalue problem, Eq.~(\ref{eq:dynmat}). 
Treating the phonon wavevector as a perturbation with respect to the zone center case, similarly to the approach of Ref.~\onlinecite{Born1996}:
\begin{equation}
\tilde{\Phi}_{\kappa\alpha,\kappa'\beta}(q) = \tilde{\Phi}_{\kappa\alpha,\kappa'\beta}(0)+\sum_{\gamma}q_\gamma \left.\frac{\partial\tilde{\Phi}_{\kappa\alpha,\kappa'\beta}}{\partial q_\gamma}\right|_0 +\cdots,
\end{equation}
\begin{equation}
U_{m\mathbf{q}}=U_{m\mathbf{0}}+\sum_\gamma q_\gamma \left.\frac{\partial U_{m\mathbf{q}}}{\partial q_\gamma}\right|_0+\cdots,
\end{equation}
and
\begin{equation}
\omega^2_{m\mathbf{q}}=\omega^2_{m\mathbf{0}}+\sum_\gamma q_\gamma \left.\frac{\partial \omega^2_{m\mathbf{q}}}{\partial q_\gamma}\right|_0+\cdots
\end{equation}
At first order for acoustic modes, one then gets
\begin{equation}
M_{\mathrm{tot}}\left.\frac{\partial \omega^2_{m\mathbf{q}}}{\partial q_\gamma}\right|_0=\sum_{\kappa\alpha,\kappa'\beta}\left.\frac{\partial\tilde{\Phi}_{\kappa\alpha,\kappa'\beta}}{\partial q_\gamma}\right|_0 U_{m\mathbf{0}}(\kappa\alpha) U_{m\mathbf{0}}(\kappa'\beta),
\end{equation}
where $M_{\mathrm{tot}}=\sum_\kappa M_\kappa$ is the total mass of the system with $M_\kappa$ the mass of the nucleus $\kappa$. This first-order term automatically vanishes due to the Hermiticity of the IFCs matrix (the corresponding first derivative is also Hermitian but only with imaginary components because of Eq.~(\ref{eq:ifc}) and the fact that real-space IFCs are purely real). For sake of compactness, we use matrix representation whenever the $\kappa\alpha$ and/or $\kappa'\beta$ indices are involved, imply that the different derivatives are estimated at $\mathbf{q}=0$ and use bracket notations, i.e.,
{\small \begin{equation}
\left.\Biggl \langle \frac{\partial \mathbf{U}_{m\mathbf{q}}}{\partial q_\gamma} \right|  \tilde{\underline{\underline{\Phi}}
}\mathbf{U}_{m\mathbf{q}}\Biggr \rangle =
\sum_{\kappa\alpha,\kappa'\beta}\left.\frac{\partial U^*_{m\mathbf{q}}(\kappa\alpha)}{\partial q_\gamma}\right|_{0}\tilde{\Phi}_{\kappa\alpha,\kappa'\beta}(0)
 U_{m\mathbf{0}}(\kappa'\beta),
\end{equation}}
with $U^*_{m\mathbf{q}}(\kappa\alpha)$ the complex conjugate of $U_{m\mathbf{q}}(\kappa\alpha)$.
Then the second order term can be written as
\begin{multline}  M_{\mathrm{tot}}\left.\frac{\partial^2 \omega^2_{m\mathbf{q}}}{\partial q_\gamma \partial q_\delta}\right|_0= 2M_{\mathrm{tot}}\left.\frac{\partial \omega_{m\mathbf{q}}}{\partial q_\gamma}\right|_0 \left.\frac{\partial \omega_{m\mathbf{q}}}{\partial q_\delta}\right|_0 =\\ \left. \Biggl \langle\frac{\partial \mathbf{U}_{m\mathbf{q}}}{\partial q_\gamma} \right|  \frac{\partial \tilde{\underline{\underline{\Phi}}}}{\partial q_\delta}
\mathbf{U}_{m\mathbf{0}}\Biggr \rangle+\left. \Biggl \langle \mathbf{U}_{m\mathbf{0}}\right| \frac{\partial \tilde{\underline{\underline{\Phi}}}}{\partial q_\delta}
 \frac{\partial \mathbf{U}_{m\mathbf{q}}}{\partial q_\gamma} \Biggr \rangle\\
 +\left. \Biggl \langle \mathbf{U}_{m\mathbf{0}}\right|\frac{\partial^2 \tilde{\underline{\underline{\Phi}}}}{\partial q_\gamma \partial q_\delta}\mathbf{U}_{m\mathbf{0}} \Biggr \rangle. \label{eq:domegadq}
\end{multline}
As one can see, both the first and second derivatives of the IFCs are involved at the second order of the perturbation theory and dictate the linearity of the phonon dispersion.
In order to get a purely quadratic mode for the flexural mode, all the corresponding terms must vanish or cancel each others. It is possible to simplify further those expressions using the first-order term~\cite{Born1996}:
\begin{equation}
\frac{\partial \mathbf{U}_{m\mathbf{q}}}{\partial q_\delta}= -\underline{\underline{\tilde{\Phi}}}^{+}(\mathbf{0})
\frac{\partial \tilde{\underline{\underline{\Phi}}}}{\partial q_\delta} \mathbf{U}_{m\mathbf{0}}
\end{equation}
with $\underline{\underline{\tilde{\Phi}}}^{+}(\mathbf{0})$ the pseudoinverse of the zone-center IFCs (the IFCs matrix having three zero eigenvalues, it is not directly invertible, explaining the use of a pseudoinverse here). Moving to real space using Eq.~(\ref{eq:ifc}), we can write equivalently:
\begin{equation}
\tilde{\Phi}_{\kappa\alpha,\kappa'\beta}(\mathbf{q})=\frac{1}{N}\sum_{\mathbf{a},\mathbf{b}} \Phi_{\kappa\alpha,\kappa'\beta}(\mathbf{a},\mathbf{b})e^{2\pi j\mathbf{q}.(R^{\mathbf{b}}_{\kappa'}-R^{\mathbf{a}}_{\kappa})},
\end{equation}
with $N$ an integer corresponding to the number of the summed unit cells. Then, considering the periodicity of the lattice:

\begin{multline}
\left. \Biggl \langle \mathbf{U}_{m\mathbf{0}}\right|\frac{\partial^2 \tilde{\underline{\underline{\Phi}}}}{\partial q_\gamma \partial q_\delta}\mathbf{U}_{m\mathbf{0}} \Biggr \rangle=\frac{(2\pi)^2}{N} \sum_{\kappa\mathbf{a},\kappa'\mathbf{b}}\Phi_{\kappa\alpha,\kappa'\beta}(\mathbf{a},\mathbf{b})\\\left[R^{\mathbf{a}}_{\kappa\gamma}R^{\mathbf{b}}_{\kappa'\delta}+R^{\mathbf{a}}_{\kappa\delta}R^{\mathbf{b}}_{\kappa'\gamma}\right]. 
\end{multline}
Let us then hypothesize the material is bidimensional, with the non-periodic direction along z. Then using Eq.~(\ref{eq:2dcase}) and the fact that $R^\mathbf{b}_{\kappa z}=R^\mathbf{0}_{\kappa z}$, we can write:
\begin{multline}
\left. \Biggl \langle \mathbf{U}_{m\mathbf{0}}\right| \frac{\partial^2 \tilde{\underline{\underline{\Phi}}}}{\partial q_\gamma \partial q_\delta}\mathbf{U}_{m\mathbf{0}} \Biggr \rangle=4\pi^2
\sum_{\kappa,\kappa'} \tilde{\Phi}_{\kappa \gamma,\kappa'\delta}(\mathbf{0}) R^{\mathbf{0}}_{\kappa z} R^{\mathbf{0}}_{\kappa'z}\\
+4\pi^2
\sum_{\kappa,\kappa'} \tilde{\Phi}_{\kappa \delta,\kappa'\gamma}(\mathbf{0}) R^{\mathbf{0}}_{\kappa z} R^{\mathbf{0}}_{\kappa'z}.\label{eq:pertd2dq}\\ 
\end{multline}
Similarly, we can write using rotational invariance:
\begin{equation}
\left.\frac{\partial \mathbf{U}_{m\mathbf{q}}}{\partial q_\delta}\right|_0(\kappa'z)= 2\pi jR^{\mathbf{0}}_{\kappa'z},
\end{equation}
and applying rotational invariance on the other side of the following bracket:
\begin{equation}
\left. \Biggl \langle \frac{\partial \mathbf{U}_{m\mathbf{q}}}{\partial q_\gamma}\right| \frac{\partial \tilde{\underline{\underline{\Phi}}}}{\partial q_\delta }\mathbf{U}_{m\mathbf{0}} \Biggr \rangle= -(2\pi)^2\sum_{\kappa \kappa'} R^{\mathbf{0}}_{\kappa z}R^{\mathbf{0}}_{\kappa' z} \tilde{\Phi}_{\kappa\gamma\kappa'\delta}(\mathbf{0}). \label{eq:pertdudcdq}
\end{equation}
The complementary term in Eq.~(\ref{eq:domegadq}) gives the same value with the permutation of $\gamma$ and $\delta$ indices. Combining Eqs.~\ref{eq:domegadq}, \ref{eq:pertd2dq} and \ref{eq:pertdudcdq}, together, we thus find that the different terms cancel each other and that the linear dispersion of the flexural modes must vanish. We recover the same findings than in previous works~\cite{Croy2020,Lin2022} without having to introduce strain deformation.

\subsection{Analytically solvable 2D-model exhibiting a flexural mode with quadratic dispersion}
\label{sec:2Dmodel}

In complement to the previous appendix, in the present one, 
we propose an analytically solvable 2D-model,
made of two parallel monoatomic chains of atoms, that has a flexural mode with 
quadratic dispersion. It is translationally and rotationally invariant, moreover only with short-range IFCs.
Also, it can sustain a shear stress that would shift one chain with respect to the other along the chain axis. Thus such a resistance to shear stress does not
prevent the flexural mode to have a quadratic dispersion, in agreement with the results
in the litterature\cite{Croy2020}, and in the previous appendix.

{Each monoatomic chain of atom has
directional (longitudinal) springs connecting the nearest-neighbor atoms, and bending springs constraining the angle between nearest-neighbor pairs of atoms, see
Fig.~\ref{fig:monoatomic_chain}.
Then, in order to connect the two parallel chains, two sets of interchain springs are considered. The first set simply creates a restoring force to keep the interchain distance constant. The second one not only keep the interchain distance constant, but also creates a restoring force with respect to shear strain between the chains.

The position of atoms in the two chains are confined in the $x$-$z$ plane: we do not consider the $y$ direction. The goal is to understand the flexural physics in this $x$-$z$ plane. Generalization to include replicated chains in the $y$ direction, and control of $y$ atomic displacements by means of adequate out-of-plane springs would not yield additional physics, and would unnecessarily complicate the argument.

The IFCs from longitudinal and bending springs are translationally and rotationally invariant, for both intrachain and interchain springs, and are short range. This is also true for their sum.
%

%
\begin{figure}
\includegraphics[width=0.48\textwidth]{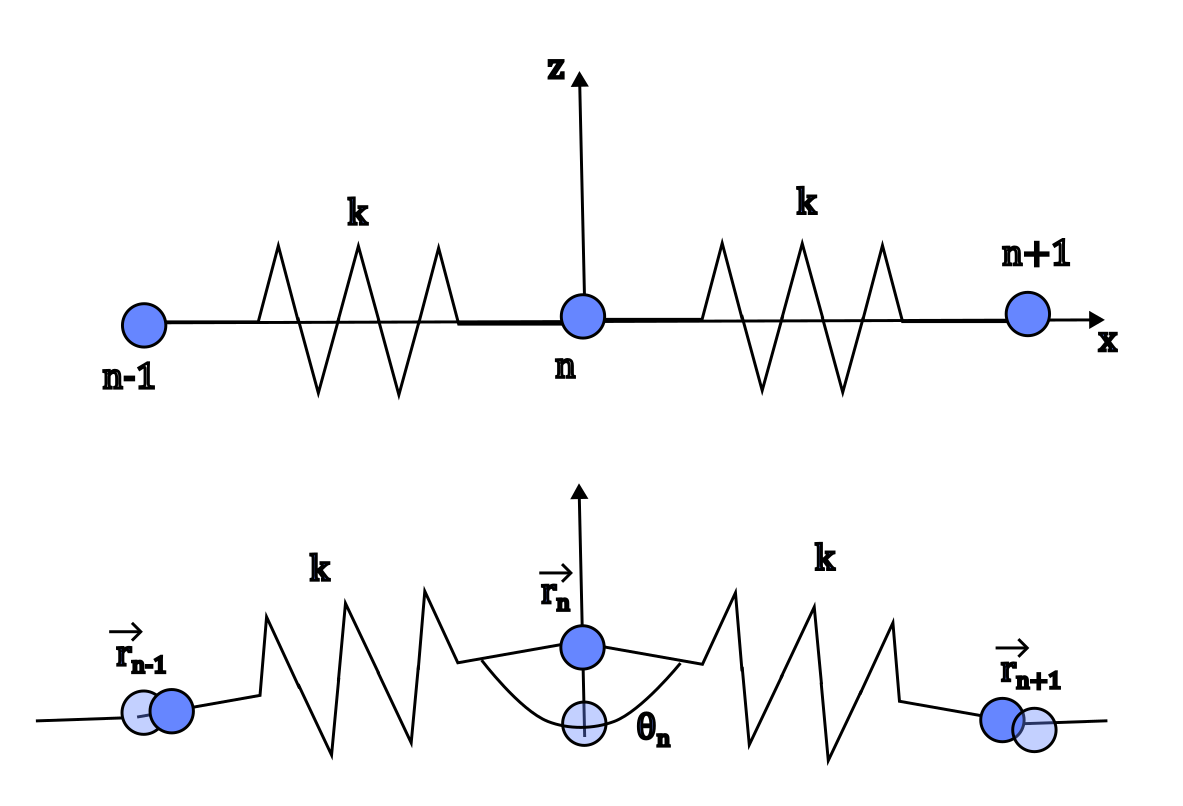}
\caption{Schematic representation of a bi-dimensional chain of atoms, which includes both a directional spring component, related to the change of interatomic distance between nearest neighbors, but also an angular spring component, related to the change of angle $\theta$ when the atoms are moved without changing the bond lengths between them. }\label{fig:monoatomic_chain}
\end{figure}

We treat first the isolated monoatomic chain.
The two contributions to the energy write 
\begin{multline}
E_1= \sum_n
\Big\{
\frac{k_\mathrm{intra}}{2}(|\mathbf{r}_{n+1}-\mathbf{r}_n|-a)^2
\\
+\frac{k_\theta}{2}
\Big(
(\mathbf{r}_{n+1}-\mathbf{r}_{n})
\times
(\mathbf{r}_{n}-\mathbf{r}_{n-1})]^2
\Big)
\Big\}
\\
\end{multline}
with $k_\mathrm{intra}$ and $k_\theta$ the radial and angular spring constants, $\mathbf{r}_n=(x_n,z_n)$ the coordinate of the atom $n$, $a$ the initial (relaxed) lattice constant, and $\times$ is a vector product.
The rotational and translational invariance of such energy is obvious, as only differences of atomic coordinates are present (hence the translational invariance), in the form of norms or vector products (all rotationally invariant) between nearest neighbors (short range).
Accordingly, we define a more compact notation, as follows,
\begin{equation}
\mathbf{r}_{i,j}
=
\mathbf{r}_{i}-\mathbf{r}_j
\end{equation}
giving
\begin{align}
E_1=
\sum_n 
\Big\{
&
\frac{k_\mathrm{intra}}{2} (|\mathbf{r}_{n+1,n}|-a)^2
\\
&+\frac{k_\theta}{2}(\mathbf{r}_{n+1,n} \times \mathbf{r}_{n,n-1})^2
\Big\}
\end{align}
}

The equilibrium coordinates of the atoms are
$\mathbf{r}_{n,\mathrm{eq}}=na\mathbf{e}_x$, where 
$\mathbf{e}_x$ is the unit vector along the chain axis.
Keeping only terms quadratic in the displacements 
$\Delta \mathbf{r}_{n}$ around 
$\mathbf{r}_{n,\mathrm{eq}}$,
one finds:
\begin{align}
E_1=
\sum_n 
\Big\{
&
\frac{k_\mathrm{intra}}{2}(x_{n,n-1}-a)^2
\nonumber \\ &
+\frac{k_\theta}{2}a^2
\left(z_{n+1,n}-z_{n,n-1}\right)^2
\Big\}
+\mathcal{O}\big((\Delta \mathbf{r}_{n})^3\big)
\end{align}

In the small displacement regime, using Bloch's theorem (or equivalently, computing the IFCs then Fourier transforming), one deduces that the eigenfrequencies $\omega$ are such that the following determinant must vanish:
\begin{equation}
\begin{vmatrix}
4k_{\mathrm{intra}}\sin^2\left(\frac{qa}{2}\right)
-M\omega^2
& 0
\\
0 & 16a^2k_\theta\sin^4\left(\frac{qa}{2}\right)
-M\omega^2
\end{vmatrix},
\end{equation}
where $M$ is the mass of the atoms and $q$ the phonon wavevector (propagation along the $x$ direction).

The upper left element yields a linear dispersion of the $x$-polarized phonon mode at small wavevector :
\begin{equation}
\label{eq:omega_x}
\omega_x(q)=2\sqrt{\frac{k_{\mathrm{intra}}}{M}}\left|\sin\left(\frac{qa}{2}\right)\right|.
\end{equation} 
\\
The lower right element yields a quadratic dispersion for the $z$-polarized phonon mode at small wavevector,
\begin{equation}
\label{eq:omega_z}
\omega_z(q)=4a\sqrt{\frac{k_\theta}{M}}\sin^2\left(\frac{qa}{2}\right).
\end{equation}
This mode is indeed flexural, exhibiting quadratic dispersion for small values of $q$.

Two monoatomic chains are then considered together.
The coordinates of the atoms acquire a subscript ``0" or ``1" depending on the chain to which they belong.
In their undistorted geometry, they are are parallel and separated perpendicularly by a distance $a$ (the same as the intrachain interatomic distance, for sake of simplicity).
By convention, the axis $x$ is aligned with the chain direction, like in the monoatomic chain case. The second chain is displaced with respect to the first chain by a vector aligned along the $z$ axis.
Explicitly, the equilibrium coordinates of the atoms are
$\mathbf{r}_{0n}=na\mathbf{e}_x$, 
and
$\mathbf{r}_{1n}=na\mathbf{e}_x+a\mathbf{e}_z$
where 
$\mathbf{e}_z$ is the unit vector perpendicular to the chain axes, connecting the two chains.
The first set of springs, perpendicular to the chain axes, with spring constant $k_{\perp}$, 
generate a force when the intrachain distance
change from its equilibrium value $a$, but do not generate a reaction to shear strain between the chains.
The second set of springs connects diagonally 
pairs of atoms that belong to different chain, with a difference of one for the $x$ index,
with spring constant $k_{\mathrm{d}}$ (for diagonal), and equilibrium length $\sqrt{2}a$. They also tend to pin the interchain distance to $a$, like the first set of springs, albeit, collectively, they also create a resistance to shear stress.
\\
Thus, there are four contributions to the energy,
\begin{widetext}
\begin{align}
E_{2,\mathrm{\perp}}=
&\sum_n \sum_{\sigma=0,1}
\Big\{
\frac{k_\mathrm{intra}}{2} 
(|\mathbf{r}_{\sigma n+1,\sigma n}|-a)^2
+\frac{k_\theta}{2}
(\mathbf{r}_{\sigma n+1,\sigma n}
\times \mathbf{r}_{\sigma n,\sigma n-1})^2
\Big\}
\nonumber
\\+
&\sum_n
\frac{k_\perp}{2}
(|\mathbf{r}_{1n,0n}|-a)^2
+\sum_n \sum_{\sigma=0,1}
\frac{k_{\mathrm{d}}}{2}
(|\mathbf{r}_{(1-\sigma)n+1,\sigma n}|-\sqrt{2}a)^2
\end{align}
Similarly to the monoatomic case, one obtains  
the eigenfrequencies $\omega$ when the following determinant vanishes:
\begin{equation}
\begin{vmatrix}
\label{eq:determinant_4}
M\omega_x^2(q)+k_{\mathrm{d}}-M\omega^2
& 0 
& -k_{\mathrm{d}}\cos(q) 
& -ik_{\mathrm{d}}\sin(q)
\\
0 & \omega_z^2(q)+k_{\perp}+k_{\mathrm{d}}
-M\omega^2
 & ik_{\mathrm{d}}\sin(q) 
& -k_{\mathrm{d}}\cos(q) -k_{\perp}
\\
-k_{\mathrm{d}}\cos(q) 
& ik_{\mathrm{d}}\sin(q) & \omega_x^2(q)+k_{\mathrm{d}}
-M\omega^2
 & 0
\\
ik_{\mathrm{d}}\sin(q) 
& -k_{\mathrm{d}}\cos(q)-k_{\perp}
& 0
&
M\omega_x^2(q)+k_{\perp}+k_{\mathrm{d}}-M\omega^2
\end{vmatrix}=0,
\end{equation}
where $\omega_x(q)$ and $\omega_z^2(q)$
are defined by Eqs.(\ref{eq:omega_x}) and
(\ref{eq:omega_z}).
The first line/row of determinant 
Eq.~(\ref{eq:determinant_4})
corresponds to movement of the atoms in the first chain in the $x$ direction, the second line/row corresponds to atoms in the first chain moving in the $z$ direction, while the third and fourth lines/rows
refer similarly to the second chain.
Due to the symmetry of the system, modes can be classified as being ``even" or ``odd" with respect to the mirror symmetry that maps one chain on the other, and this is irrespective of the value of the wavevector (that is invariant under such symmetry). The 4x4 matrix can thus be block-diagonalized. For the even modes and the odd modes, respectively, the following determinants must vanish, respectively,
\begin{equation}
\begin{vmatrix}
M\omega_x^2(q)+k_{\mathrm{d}}(1-\cos(q))-M\omega^2 
& ik_{\mathrm{d}}\sin(q)
\\
-ik_{\mathrm{d}}\sin(q) 
&
M\omega_z^2(q)+k_{\perp}(1+\cos(q))+2k_{\mathrm{d}}-M\omega^2
\end{vmatrix}=0,
\end{equation}
\begin{equation}
\label{eq:odd_determinant}
\begin{vmatrix}
M\omega_x^2(q)+k_{\mathrm{d}}(1+\cos(q))-M\omega^2 
& ik_{\mathrm{d}}\sin(q)
\\
-ik_{\mathrm{d}}\sin(q) 
&
M\omega_z^2(q)+k_{\perp}(1-\cos(q))-M\omega^2
\end{vmatrix}=0.
\end{equation}
The flexural modes are odd under the above-mentioned mirror symmetry. The corresponding determinant 
Eq.~(\ref{eq:odd_determinant}) 
vanishes for 
\begin{equation}
M\omega^2=M(\frac{\omega_x^2+\omega_z^2}{2})+k_\textrm{d}
\pm
\sqrt{
\Bigg(
M(\frac{\omega_x^2+\omega_z^2}{2})+k_\textrm{d}\Bigg)^2
-\Big(
M\omega_x^2 M\omega_z^2
- 
M\omega_x^2 k_\textrm{d}(1-\cos(q))
-
M\omega_z^2 k_\textrm{d}(1+\cos(q))
\Big),
}
\end{equation}
the minus sign corresponding to the flexural mode. 
It is easy to check that the second term
between parentheses within the square root behaves like $q^4$, because $\omega_z^2$ behaves like $q^4$, while 
$\omega_x^2$ and $1-c(q)$ behave each like $q^2$. 
For the minus sign solution, this yields an overall $M\omega^2 \propto q^4$ behavior. In consequence, the flexural mode for this two parallel monoatomic chains model shows a quadratic dispersion for the flexural mode close to the zone-center, with an altered curvature compared to the single atomic chain.
\end{widetext}

\subsection{Convergence studies}

\begin{figure}[htp!]
\includegraphics[width=0.48\textwidth]{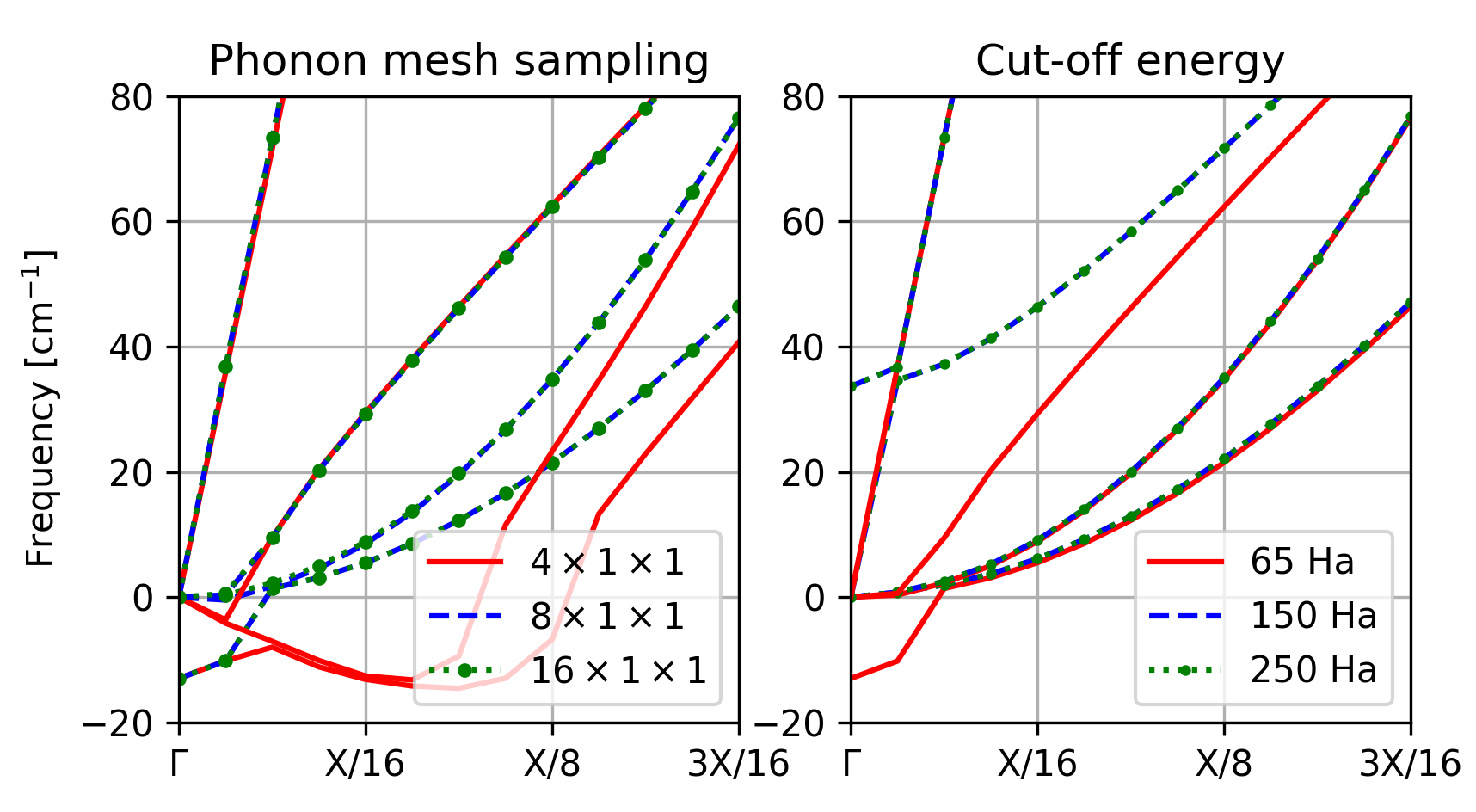}
\caption{Convergence of the acoustic modes in the polyethylene chain with respect to (left) computed q-wavevectors mesh (solid red: 4 q-points, dash blue: 8 q-points, dotted green: 16 q-points) and (right) plane-wave cut-off energy.}
\label{fig:1D_qptconv}
\end{figure}

In this Appendix, we study the convergence with respect to two numerical parameters (phonon Brillouin Zone sampling and electronic planewave cut-off energy) for the materials under study. Results for the polyethylene chain are shown in Fig.~\ref{fig:1D_qptconv}, with only translational invariance imposed.
We use a fixed 16x1x1 wavevector grid for the
electronic wavefunctions.

On the left, using a 65~Ha cut-off energy, the results for three phonon wavevector grids are compared. It is seen that the 
8$\times$1$\times$1 and 16$\times$1$\times$1 grids gives nearly indistinguishable results. 
However, for all grids, one of the acoustic frequencies obtained at the $\Gamma$ point does not vanish.
Fixing the phonon wavevector grid
to 8$\times$1$\times$1, the electronic planewave cut-off energy is then increased, as shown on the right. No incorrect acoustic frequencies are present for the 150 Ha or 250 Ha cut-off energies.

This indicates that the long-range electrostatics are not problematic for this 1D system. 
In terms of cut-off energies, 150~Ha is required for polyethylene with HGH pseudopotentials. The rotational mode of the chain is particularly sensitive to the cut-off energy, but this sensitivity is nullified after the imposition of the rotational invariance.

\begin{figure*}[htp!]
\includegraphics[width=0.7\textwidth]{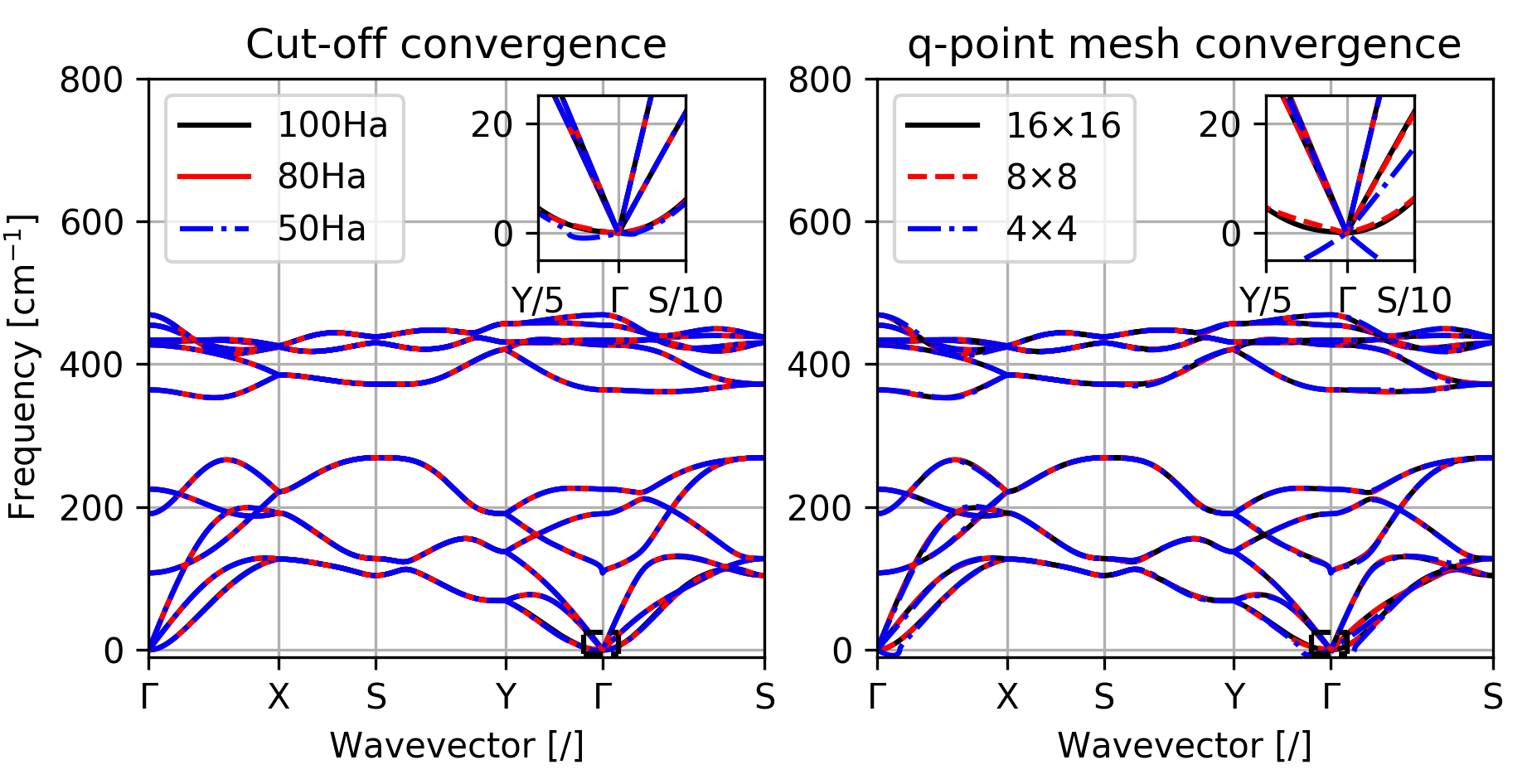}
\caption{Convergence of the acoustic modes in phosphorene for HGH pseudopotentials with increasing (left) cut-off energy and (right)  computed phonon wavevector mesh sampling.}
\label{fig:phospho_ecut}
\end{figure*}

We did the same exercise for phosphorene. The convergence studies for HGH pseudopotentials with respect to the phonon wavevector mesh density and to the cut-off energy are shown in Fig.~\ref{fig:phospho_ecut}. As one can see, the only problematic region is the acoustic modes along the $\Gamma \rightarrow Y$ and $\Gamma\rightarrow S$ high-symmetry lines. The flexural mode is particularly challenging to converge with respect to the cut-off energy, with some oscillatory behavior. In contrast, the ONCVPSP pseudopotentials only require 30~Ha to reach convergence (not shown here).  

\begin{figure}[h]
\includegraphics[width=0.39\textwidth]{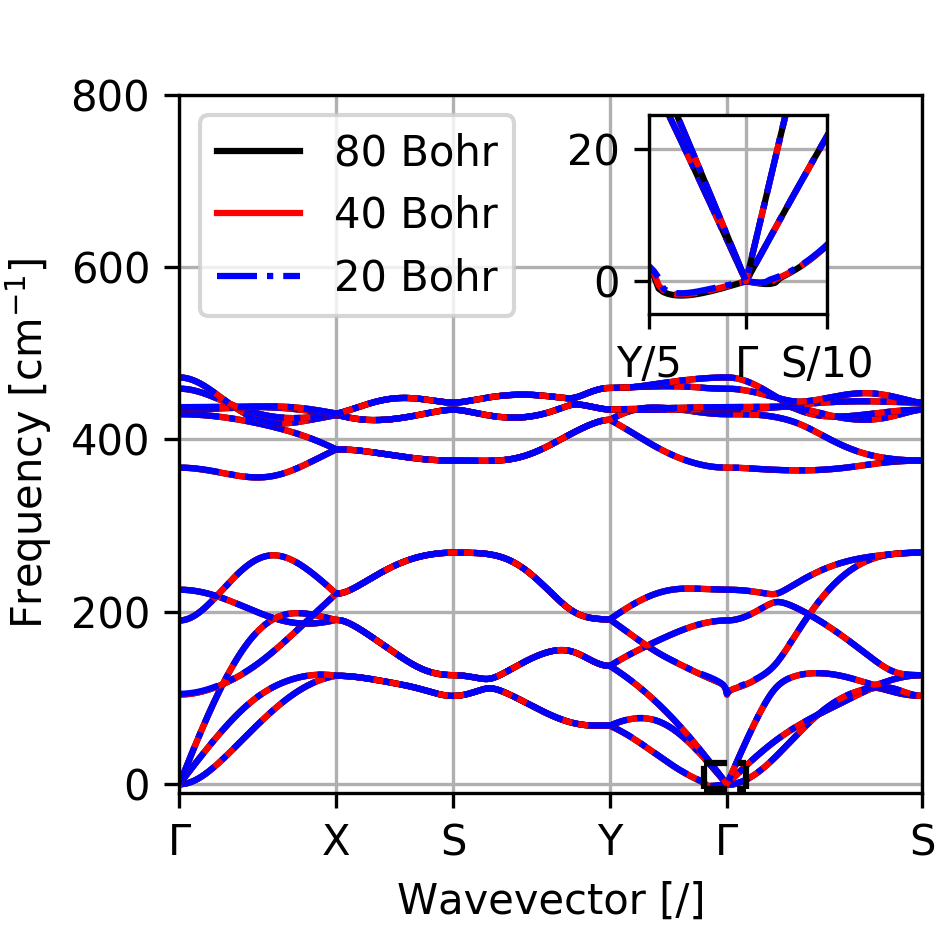}
\caption{Variation of the phonon band structure of phosphorene with the size of the unit cell in the out-of-plane parameter used to isolate the layer for ONCVPSP pseudopotentials.}
\label{fig:phospho_vac}
\end{figure}

Then, we investigated the impact of the vacuum thickness on the phosphorene phonon band structure, going from 20~Bohr to 80~Bohr. 
We show the results for the ONCVPSP pseudopotentials with a 30~Ha cut-off energy and only translational invariance imposed in Fig.~\ref{fig:phospho_vac}.
From this plot, we deduce that the negative mode for the ONCVPSP is not a consequence of the interactions between the images, but directly from the breaking of the translational and rotational invariance.
The ONCVPSP pseudopotentials only require 30~Ha to reach convergence (not shown here).  

\bibliography{Biblio.bib}

@article{Hartwigsen1998,
  title = {Relativistic separable dual-space Gaussian pseudopotentials from H to Rn},
  author = {Hartwigsen, C. and Goedecker, S. and Hutter, J.},
  journal = {Phys. Rev. B},
  volume = {58},
  issue = {7},
  pages = {3641--3662},
  numpages = {0},
  year = {1998},
  month = {Aug},
  publisher = {American Physical Society},
  doi = {10.1103/PhysRevB.58.3641},
  url = {https://link.aps.org/doi/10.1103/PhysRevB.58.3641}
}

@article{Croy2020,
doi = {10.1088/2515-7639/ab8271},
url = {https://doi.org/10.1088/2515-7639/ab8271},
year = {2020},
month = {jun},
publisher = {IOP Publishing},
volume = {3},
number = {2},
pages = {02LT03},
author = {Croy, Alexander},
title = {Bending rigidities and universality of flexural modes in 2D crystals},
journal = {Journal of Physics: Materials},
}

@book{Born1996,
    author = {Born, Max and Huang, Kun},
    title = {Dynamical Theory Of Crystal Lattices},
    publisher = {Oxford University Press},
    year = {1996},
    month = {08},
    isbn = {9780192670083},
    doi = {10.1093/oso/9780192670083.001.0001},
    url = {https://doi.org/10.1093/oso/9780192670083.001.0001},
}

@article{Romero2020,
author = {Romero,Aldo H.  and Allan,Douglas C.  and Amadon,Bernard  and Antonius,Gabriel  and Applencourt,Thomas  and Baguet,Lucas  and Bieder,Jordan  and Bottin,François  and Bouchet,Johann  and Bousquet,Eric  and Bruneval,Fabien  and Brunin,Guillaume  and Caliste,Damien  and Côté,Michel  and Denier,Jules  and Dreyer,Cyrus  and Ghosez,Philippe  and Giantomassi,Matteo  and Gillet,Yannick  and Gingras,Olivier  and Hamann,Donald R.  and Hautier,Geoffroy  and Jollet,François  and Jomard,Gérald  and Martin,Alexandre  and Miranda,Henrique P. C.  and Naccarato,Francesco  and Petretto,Guido  and Pike,Nicholas A.  and Planes,Valentin  and Prokhorenko,Sergei  and Rangel,Tonatiuh  and Ricci,Fabio  and Rignanese,Gian-Marco  and Royo,Miquel  and Stengel,Massimiliano  and Torrent,Marc  and van Setten,Michiel J.  and Van Troeye,Benoit  and Verstraete,Matthieu J.  and Wiktor,Julia  and Zwanziger,Josef W.  and Gonze,Xavier },
title = {ABINIT: Overview and focus on selected capabilities},
journal = {The Journal of Chemical Physics},
volume = {152},
number = {12},
pages = {124102},
year = {2020},
doi = {10.1063/1.5144261},

eprint = { 
        https://doi.org/10.1063/1.5144261
}
}

@article{Hamann2013,
  title = {Optimized norm-conserving Vanderbilt pseudopotentials},
  author = {Hamann, D. R.},
  journal = {Phys. Rev. B},
  volume = {88},
  pages = {085117},
  numpages = {10},
  year = {2013},
  doi = {10.1103/PhysRevB.88.085117},
}

@article{pseudodojo2018,
title = {The PseudoDojo: Training and grading a 85 element optimized norm-conserving pseudopotential table},
journal = {Computer Physics Communications},
volume = {226},
pages = {39-54},
year = {2018},
issn = {0010-4655},
doi = {https://doi.org/10.1016/j.cpc.2018.01.012},
url = {https://www.sciencedirect.com/science/article/pii/S0010465518300250},
author = {M.J. {van Setten} and M. Giantomassi and E. Bousquet and M.J. Verstraete and D.R. Hamann and X. Gonze and G.-M. Rignanese},
}

@article{Lejaeghere2016,
author = {Kurt Lejaeghere  and Gustav Bihlmayer  and Torbjörn Björkman  and Peter Blaha  and Stefan Blügel  and Volker Blum  and Damien Caliste  and Ivano E. Castelli  and Stewart J. Clark  and Andrea Dal Corso  and Stefano de Gironcoli  and Thierry Deutsch  and John Kay Dewhurst  and Igor Di Marco  and Claudia Draxl  and Marcin Dułak  and Olle Eriksson  and José A. Flores-Livas  and Kevin F. Garrity  and Luigi Genovese  and Paolo Giannozzi  and Matteo Giantomassi  and Stefan Goedecker  and Xavier Gonze  and Oscar Grånäs  and E. K. U. Gross  and Andris Gulans  and François Gygi  and D. R. Hamann  and Phil J. Hasnip  and N. A. W. Holzwarth  and Diana Iuşan  and Dominik B. Jochym  and François Jollet  and Daniel Jones  and Georg Kresse  and Klaus Koepernik  and Emine Küçükbenli  and Yaroslav O. Kvashnin  and Inka L. M. Locht  and Sven Lubeck  and Martijn Marsman  and Nicola Marzari  and Ulrike Nitzsche  and Lars Nordström  and Taisuke Ozaki  and Lorenzo Paulatto  and Chris J. Pickard  and Ward Poelmans  and Matt I. J. Probert  and Keith Refson  and Manuel Richter  and Gian-Marco Rignanese  and Santanu Saha  and Matthias Scheffler  and Martin Schlipf  and Karlheinz Schwarz  and Sangeeta Sharma  and Francesca Tavazza  and Patrik Thunström  and Alexandre Tkatchenko  and Marc Torrent  and David Vanderbilt  and Michiel J. van Setten  and Veronique Van Speybroeck  and John M. Wills  and Jonathan R. Yates  and Guo-Xu Zhang  and Stefaan Cottenier },
title = {Reproducibility in density functional theory calculations of solids},
journal = {Science},
volume = {351},
number = {6280},
pages = {aad3000},
year = {2016},
doi = {10.1126/science.aad3000},
URL = {https://www.science.org/doi/abs/10.1126/science.aad3000},
eprint = {https://www.science.org/doi/pdf/10.1126/science.aad3000},
}

@article{VanTroeye2016,
  title = {Interatomic force constants including the DFT-D dispersion contribution},
  author = {Van Troeye, Benoit and Torrent, Marc and Gonze, Xavier},
  journal = {Phys. Rev. B},
  volume = {93},
  issue = {14},
  pages = {144304},
  numpages = {9},
  year = {2016},
  month = {Apr},
  publisher = {American Physical Society},
  doi = {10.1103/PhysRevB.93.144304},
  url = {https://link.aps.org/doi/10.1103/PhysRevB.93.144304}
}

@article{Sohier2016,
  title = {Two-dimensional Fr\"ohlich interaction in transition-metal dichalcogenide monolayers: Theoretical modeling and first-principles calculations},
  author = {Sohier, Thibault and Calandra, Matteo and Mauri, Francesco},
  journal = {Phys. Rev. B},
  volume = {94},
  issue = {8},
  pages = {085415},
  numpages = {13},
  year = {2016},
  month = {Aug},
  publisher = {American Physical Society},
  doi = {10.1103/PhysRevB.94.085415},
  url = {https://link.aps.org/doi/10.1103/PhysRevB.94.085415}
}

@article{Sohier2017,
	title = {Breakdown of {Optical} {Phonons}’ {Splitting} in {Two}-{Dimensional} {Materials}},
	volume = {17},
	issn = {1530-6984},
	url = {https://doi.org/10.1021/acs.nanolett.7b01090},
	doi = {10.1021/acs.nanolett.7b01090},
	number = {6},
	journal = {Nano Letters},
	author = {Sohier, Thibault and Gibertini, Marco and Calandra, Matteo and Mauri, Francesco and Marzari, Nicola},
	month = jun,
	year = {2017},
	note = {Publisher: American Chemical Society},
	pages = {3758--3763},
}

@article{Stengel2013,
  title = {Flexoelectricity from density-functional perturbation theory},
  author = {Stengel, Massimiliano},
  journal = {Phys. Rev. B},
  volume = {88},
  issue = {17},
  pages = {174106},
  numpages = {24},
  year = {2013},
  month = {Nov},
  publisher = {American Physical Society},
  doi = {10.1103/PhysRevB.88.174106},
  url = {https://link.aps.org/doi/10.1103/PhysRevB.88.174106}
}

@article{Carrete2016,
author = {Jesús Carrete and Wu Li and Lucas Lindsay and David A. Broido and Luis J. Gallego and Natalio Mingo and},
title = {Physically founded phonon dispersions of few-layer materials and the case of borophene},
journal = {Materials Research Letters},
volume = {4},
number = {4},
pages = {204--211},
year = {2016},
publisher = {Taylor \& Francis},
doi = {10.1080/21663831.2016.1174163},
URL = { https://doi.org/10.1080/21663831.2016.1174163
},
eprint = { https://doi.org/10.1080/21663831.2016.1174163
}
}

@article {Novoselov2016,
	author = {Novoselov, K. S. and Mishchenko, A. and Carvalho, A. and Castro Neto, A. H.},
	title = {2D materials and van der Waals heterostructures},
	volume = {353},
	year = {2016},
	pages = {aac9439},
	doi = {10.1126/science.aac9439},
	publisher = {American Association for the Advancement of Science},
	journal = {Science}
}

@article{Pick1970,
  title = {Microscopic Theory of Force Constants in the Adiabatic Approximation},
  author = {Pick, Robert M. and Cohen, Morrel H. and Martin, Richard M.},
  journal = {Phys. Rev. B},
  volume = {1},
  issue = {2},
  pages = {910--920},
  numpages = {0},
  year = {1970},
  month = {Jan},
  publisher = {American Physical Society},
  doi = {10.1103/PhysRevB.1.910},
  url = {https://link.aps.org/doi/10.1103/PhysRevB.1.910}
}

@article{Ankit2015,
	title = {Strongly anisotropic in-plane thermal transport in single-layer black phosphorene},
	volume = {5},
	issn = {2045-2322},
	url = {https://doi.org/10.1038/srep08501},
	doi = {10.1038/srep08501},
	number = {1},
	journal = {Scientific Reports},
	author = {Jain, Ankit and McGaughey, Alan J. H.},
	month = feb,
	year = {2015},
	pages = {8501},
}

@article{Grimme2006,
        author = {Stephan Grimme},
        title = {Semiempirical GGA-Type Density Functional Constructed
with a Long-Range Dispersion Correction},
        journal = {J. Comput. Chem.},
        volume = {27},
        pages = {1787},
        year = {2006},
        doi = {10.1002/jcc.20495},
}

@article{VanTroeye2025,
author={Benoit Van Troeye and Geoffrey Pourtois},
journal={submitted}
}

@article{Lin2023,
author = {Lin, Yu-Chuan and Torsi, Riccardo and Younas, Rehan and Hinkle, Christopher L. and Rigosi, Albert F. and Hill, Heather M. and Zhang, Kunyan and Huang, Shengxi and Shuck, Christopher E. and Chen, Chen and Lin, Yu-Hsiu and Maldonado-Lopez, Daniel and Mendoza-Cortes, Jose L. and Ferrier, John and Kar, Swastik and Nayir, Nadire and Rajabpour, Siavash and van Duin, Adri C. T. and Liu, Xiwen and Jariwala, Deep and Jiang, Jie and Shi, Jian and Mortelmans, Wouter and Jaramillo, Rafael and Lopes, Joao Marcelo J. and Engel-Herbert, Roman and Trofe, Anthony and Ignatova, Tetyana and Lee, Seng Huat and Mao, Zhiqiang and Damian, Leticia and Wang, Yuanxi and Steves, Megan A. and Knappenberger, Kenneth L. Jr. and Wang, Zhengtianye and Law, Stephanie and Bepete, George and Zhou, Da and Lin, Jiang-Xiazi and Scheurer, Mathias S. and Li, Jia and Wang, Pengjie and Yu, Guo and Wu, Sanfeng and Akinwande, Deji and Redwing, Joan M. and Terrones, Mauricio and Robinson, Joshua A.},
title = {Recent Advances in 2D Material Theory, Synthesis, Properties, and Applications},
journal = {ACS Nano},
volume = {17},
number = {11},
pages = {9694-9747},
year = {2023},
doi = {10.1021/acsnano.2c12759},
    note ={PMID: 37219929},
URL = { 
https://doi.org/10.1021/acsnano.2c12759
},
eprint = { https://doi.org/10.1021/acsnano.2c12759
}
}

@article{Lemme2022,
	title = {2D materials for future heterogeneous electronics},
	volume = {13},
	issn = {2041-1723},
	url = {https://doi.org/10.1038/s41467-022-29001-4},
	doi = {10.1038/s41467-022-29001-4},
	pages = {1392},
	number = {1},
	journal = {Nature Communications},
	shortjournal = {Nature Communications},
	author = {Lemme, Max C. and Akinwande, Deji and Huyghebaert, Cedric and Stampfer, Christoph},
	date = {2022-03-16},
        year ={2022},
}

@article{Obrien2023,
	title = {Process integration and future outlook of 2D transistors},
	volume = {14},
	issn = {2041-1723},
	url = {https://doi.org/10.1038/s41467-023-41779-5},
	doi = {10.1038/s41467-023-41779-5},
	pages = {6400},
	number = {1},
	journal = {Nature Communications},
	shortjournal = {Nature Communications},
	author = {O’Brien, Kevin P. and Naylor, Carl H. and Dorow, Chelsey and Maxey, Kirby and Penumatcha, Ashish Verma and Vyatskikh, Andrey and Zhong, Ting and Kitamura, Ande and Lee, Sudarat and Rogan, Carly and Mortelmans, Wouter and Kavrik, Mahmut Sami and Steinhardt, Rachel and Buragohain, Pratyush and Dutta, Sourav and Tronic, Tristan and Clendenning, Scott and Fischer, Paul and Putna, Ernisse S. and Radosavljevic, Marko and Metz, Matt and Avci, Uygar},
	date = {2023-10-12},
        year = {2023},
}

@article{Lin2022,
	title = {General invariance and equilibrium conditions for lattice dynamics in {1D}, {2D}, and {3D} materials},
	volume = {8},
	issn = {2057-3960},
	url = {https://doi.org/10.1038/s41524-022-00920-6},
	doi = {10.1038/s41524-022-00920-6},
    number = {1},
	journal = {npj Computational Materials},
	author = {Lin, Changpeng and Poncé, Samuel and Marzari, Nicola},
	month = nov,
	year = {2022},
	pages = {236},
}

@book{Landau2012,
  title={Theory of elasticity. Volume 7 of the L.D. Landau Course of Theoretical Physics.},
  author={Pitaevskii, LP and Kosevich, Arnolʹd Markovich and Lifshitz, Evgenii Mikhailovich},
  year={2012},
  publisher={Elsevier}
}

@article{Ponce2023,
  title = {Long-range electrostatic contribution to electron-phonon couplings and mobilities of two-dimensional and bulk materials},
  author = {Ponc\'e, Samuel and Royo, Miquel and Stengel, Massimiliano and Marzari, Nicola and Gibertini, Marco},
  journal = {Phys. Rev. B},
  volume = {107},
  issue = {15},
  pages = {155424},
  numpages = {30},
  year = {2023},
  month = {Apr},
  publisher = {American Physical Society},
  doi = {10.1103/PhysRevB.107.155424},
  url = {https://link.aps.org/doi/10.1103/PhysRevB.107.155424}
}

@article{Brunin2020,
  title = {Electron-Phonon beyond Fr\"ohlich: Dynamical Quadrupoles in Polar and Covalent Solids},
  author = {Brunin, Guillaume and Miranda, Henrique Pereira Coutada and Giantomassi, Matteo and Royo, Miquel and Stengel, Massimiliano and Verstraete, Matthieu J. and Gonze, Xavier and Rignanese, Gian-Marco and Hautier, Geoffroy},
  journal = {Phys. Rev. Lett.},
  volume = {125},
  issue = {13},
  pages = {136601},
  numpages = {6},
  year = {2020},
  month = {Sep},
  publisher = {American Physical Society},
  doi = {10.1103/PhysRevLett.125.136601},
  url = {https://link.aps.org/doi/10.1103/PhysRevLett.125.136601}
}

@article{Royo2020b,
  title = {Using High Multipolar Orders to Reconstruct the Sound Velocity in Piezoelectrics from Lattice Dynamics},
  author = {Royo, Miquel and Hahn, Konstanze R. and Stengel, Massimiliano},
  journal = {Phys. Rev. Lett.},
  volume = {125},
  issue = {21},
  pages = {217602},
  numpages = {7},
  year = {2020},
  month = {Nov},
  publisher = {American Physical Society},
  doi = {10.1103/PhysRevLett.125.217602},
  url = {https://link.aps.org/doi/10.1103/PhysRevLett.125.217602}
}

@article{Royo2019,
  title = {First-Principles Theory of Spatial Dispersion: Dynamical Quadrupoles and Flexoelectricity},
  author = {Royo, Miquel and Stengel, Massimiliano},
  journal = {Phys. Rev. X},
  volume = {9},
  issue = {2},
  pages = {021050},
  numpages = {22},
  year = {2019},
  month = {Jun},
  publisher = {American Physical Society},
  doi = {10.1103/PhysRevX.9.021050},
  url = {https://link.aps.org/doi/10.1103/PhysRevX.9.021050}
}

@article{Sohier2017b,
  title = {Density functional perturbation theory for gated two-dimensional heterostructures: Theoretical developments and application to flexural phonons in graphene},
  author = {Sohier, Thibault and Calandra, Matteo and Mauri, Francesco},
  journal = {Phys. Rev. B},
  volume = {96},
  issue = {7},
  pages = {075448},
  numpages = {21},
  year = {2017},
  month = {Aug},
  publisher = {American Physical Society},
  doi = {10.1103/PhysRevB.96.075448},
  url = {https://link.aps.org/doi/10.1103/PhysRevB.96.075448}
}

@article{Rivano2023,
	title = {Infrared-active phonons in one-dimensional materials and their spectroscopic signatures},
	volume = {9},
	issn = {2057-3960},
	url = {https://doi.org/10.1038/s41524-023-01140-2},
	doi = {10.1038/s41524-023-01140-2},
	number = {1},
	journal = {npj Computational Materials},
	author = {Rivano, Norma and Marzari, Nicola and Sohier, Thibault},
	month = oct,
	year = {2023},
	pages = {194},
}

@article{Rivano2024,
  title = {Density functional perturbation theory for one-dimensional systems: Implementation and relevance for phonons and electron-phonon interactions},
  author = {Rivano, Norma and Marzari, Nicola and Sohier, Thibault},
  journal = {Phys. Rev. B},
  volume = {109},
  issue = {24},
  pages = {245426},
  numpages = {13},
  year = {2024},
  month = {Jun},
  publisher = {American Physical Society},
  doi = {10.1103/PhysRevB.109.245426},
  url = {https://link.aps.org/doi/10.1103/PhysRevB.109.245426}
}

@article{Royo2022,
  title = {Lattice-mediated bulk flexoelectricity from first principles},
  author = {Royo, Miquel and Stengel, Massimiliano},
  journal = {Phys. Rev. B},
  volume = {105},
  issue = {6},
  pages = {064101},
  numpages = {17},
  year = {2022},
  month = {Feb},
  publisher = {American Physical Society},
  doi = {10.1103/PhysRevB.105.064101},
  url = {https://link.aps.org/doi/10.1103/PhysRevB.105.064101}
}

@article{Royo2021,
  title = {Exact Long-Range Dielectric Screening and Interatomic Force Constants in Quasi-Two-Dimensional Crystals},
  author = {Royo, Miquel and Stengel, Massimiliano},
  journal = {Phys. Rev. X},
  volume = {11},
  issue = {4},
  pages = {041027},
  numpages = {22},
  year = {2021},
  month = {Nov},
  publisher = {American Physical Society},
  doi = {10.1103/PhysRevX.11.041027},
  url = {https://link.aps.org/doi/10.1103/PhysRevX.11.041027}
}

@article{Gonze1997,
  title = {Dynamical matrices, Born effective charges, dielectric permittivity tensors, and interatomic force constants from density-functional perturbation theory},
  author = {Gonze, Xavier and Lee, Changyol},
  journal = {Phys. Rev. B},
  volume = {55},
  issue = {16},
  pages = {10355--10368},
  numpages = {0},
  year = {1997},
  month = {Apr},
  publisher = {American Physical Society},
  doi = {10.1103/PhysRevB.55.10355},
  url = {https://link.aps.org/doi/10.1103/PhysRevB.55.10355}
}

@article{Abinit2005,
        author = {Xavier Gonze and Gian-Marco Rignanese and Matthieu Verstraete and Jean-Michel Beuken and 
        Yann Pouillon and Razvan Caracas and Francois Jollet and Marc Torrent and Gilles Zerah and Masayoshi Mikami
 and Philippe Ghosez and Marek Veithen and Jean-Yves Raty and Valerio Olevano and Fabien Bruneval and Lucia Reining
 and Rex Godby and Giovanni Onida and Donald R. Hamann and Douglas C. Allan},
        title = {A brief introduction to the ABINIT software package},
        journal = {Z. Kristallogr.},
        volume = {220},
        pages = {558},
        year = {2005},
        doi = {10.1524/zkri.220.5.558.65066},
}

@article{Abinit2009,
        author = {X. Gonze and B. Amadon and P.-M. Anglade and J.-M. Beuken and F. Bottin and P. Boulanger
        and F. Bruneval and D. Caliste and R. Caracas and M. C\^{o}t\'{e} and T. Deutsch and L. Genovese and Ph. Ghosez
        and M. Giantomassi and S. Goedecker and D.R. Hamann and P. Hermet and F. Jollet
        and G. Jomard and S. Leroux and M. Mancini and S. Mazevet and M.J.T. Oliveira and b, G. Onida and
        Y. Pouillon and T. Rangela and G.-M. Rignanese and D. Sangalli and n, R. Shaltaf and M. Torrent and
        M.J. Verstraete and G. Zerah and J.W. Zwanziger},
        title = {ABINIT: First-principles approach to material and nanosystem properties},
        journal = {Comput. Phys. Commun.},
        volume = {180},
        pages = {2582},
        doi = {10.1016/j.cpc.2009.07.007},
        year = {2009}
}

@article{Gonze2016,
title = "Recent developments in the \{ABINIT\} software package ",
journal = "Comp. Phys. Comm. ",
volume = "205",
pages = "106 - 131",
year = "2016",
note = "",
doi = "10.1016/j.cpc.2016.04.003",
author = "X. Gonze and F. Jollet and F. Abreu Araujo and D. Adams and B. Amadon and T. Applencourt and C. Audouze and J.-M. Beuken and J. Bieder and A. Bokhanchuk and E. Bousquet and F. Bruneval and D. Caliste and M. Côté and F. Dahm and F. Da Pieve and M. Delaveau and M. Di Gennaro and B. Dorado and C. Espejo and G. Geneste and L. Genovese and A. Gerossier and M. Giantomassi and Y. Gillet and D.R. Hamann and L. He and G. Jomard and J. Laflamme Janssen and S. Le Roux and A. Levitt and A. Lherbier and F. Liu and I. Lukačević and A. Martin and C. Martins and M.J.T. Oliveira and S. Poncé and Y. Pouillon and T. Rangel and G.-M. Rignanese and A.H. Romero and B. Rousseau and O. Rubel and A.A. Shukri and M. Stankovski and M. Torrent and M.J. Van Setten and B. Van Troeye and M.J. Verstraete and D. Waroquiers and J. Wiktor and B. Xu and A. Zhou and J.W. Zwanziger",
}

\end{document}